\documentclass{article}
\usepackage{amsmath,amsfonts,amssymb}
\usepackage{graphicx}
\usepackage{float}
\usepackage[textwidth=0.7\paperwidth,textheight=0.75\paperheight]{geometry}
\usepackage{natbib}

\setcitestyle{comma}

\begin{document}

\title{Exponential Dispersion Models for Overdispersed 
Zero-Inflated Count Data}

\author{Shaul K. Bar-Lev$^a$ \and Ad Ridder$^b$}

\date{ 
$^{a}$\textit{\small Faculty of Tech. Man., Holon
Institute of Technology, Holon, Israel},
\texttt{\small shaulb@hit.ac.il}\\
$^b$\textit{\small 
Department of EOR, VU University, Amsterdam, Netherlands},
\texttt{\small ad.ridder@vu.nl}\\
[2ex] \today
}

\maketitle

\begin{abstract}
We consider three new classes of exponential dispersion models of discrete
probability distributions which are defined by specifying their variance
functions in their mean value parameterization. In a previous paper (Bar-Lev
and Ridder, 2020a), we have developed the framework of these classes and
proved that they have some desirable properties. Each of these classes was shown
to be overdispersed and zero inflated in ascending order, making them as
competitive statistical models for those in use in statistical modeling. In
this paper we elaborate on the computational aspects of their probability
mass functions. Furthermore, we apply these classes for fitting real data
sets having overdispersed and zero-inflated statistics. Classic models based
on Poisson or negative binomial distributions show poor fits, and therefore
many alternatives have already proposed in recent years. We execute an
extensive comparison with these other proposals, from which we may conclude
that our framework is a flexible tool that gives excellent results in all
cases. Moreover, in most cases our model gives the best fit.


\end{abstract}

\section{Introduction}
In many scientific fields, one deals with systems, experiments, or phenomena
that have random discrete outcomes. These outcomes are revealed
typically through a set of data obtained by observing the system and
counting the occurrences. The next issue is to describe statistically the
system, modeling the occurrences by a discrete random variable, or just by a
discrete probability distribution. Indeed, this is our starting point: we
consider a number of sets of different count data, each from another
application, and our goal is to fit suitable discrete distributions. In 
\citet{barlev2020a} we have developed a framework of exponential dispersion
models for count data, which resulted in three classes of parametric
families of discrete distributions presented in terms of their variance
functions. We proved that these three classes have some desirable
properties. Each of these classes was shown to be overdispersed and zero
inflated in ascending order, making them as competitive statistical models
for those in use in statistical modeling. The contribution of this paper is,
firstly, that we elaborate on the computational aspects of these
distributions, and secondly, that we use our framework for real data sets.
We shall examine the goodness-of-fit of the proposed distributions when
applied to these data sets. And we shall compare our fits with known fits
from literature. Specifically, we consider data that show (i)
overdispersion, meaning the variance is greater than the mean; and (ii)
zero-inflation, i.e., a high occurrence of zero values. We shall see that
our framework allows for very flexible modeling in all kinds of different
statistical situations, but provides a tight fit in all cases.

\subsection{Literature Review}
Modeling count data by parametric families of discrete distributions has
always been part of the statistical literature. Traditional models of count
data were based on the Poisson and the negative binomial distributions, for
instance in case of

\begin{itemize}
\item 
actuarial sciences for car insurance claims \citep{gossiaux1981},
hospitalizations \citep{klugman1998}, motor vehicle crashes \citep{lord2005};

\item 
health economics for kidneys cysts \citep{chan2009}, Thalassemia
disease among children \citep{zafakali2013}, bovine tuberculosis cases 
\citep{coly2016};

\item 
psychology and behavioral sciences for cigarette smoking 
\citep{siddiqui1999}, alcohol drinking \citep{armeli2005}, prison incidents 
\citep{walters2007}.
\end{itemize}

The advantages of the Poisson and negative binomial distributions are their
easy computations and parameter estimation. The main disadvantage of the
Poisson distribution is its variance being equal to its mean. Thus, this
distribution does not fit the data properly in case of overdispersion. The
negative binomial distribution is an improvement since it can model
overdisperion, however it could give poor fits to data with an excessive
number of zeros \citep{coly2016}.

In view of this, many other statistical models have been proposed and
studied. An abundance of studies of new models can be observed in recent
years, the reason being the availability of many diverse data sources, but
more importantly, the advances in computational techniques. Commonly, the
purpose is to handle overdispersion, or to handle zero-inflation; less
frequently to handle both.

To name a few concerning handling overdispersion which are relevant for our
study, generalized negative binomial \citep{jain1971}, Poisson-inverse
Gaussian distribution \citep{willmot1987}, strict arcsine distribution 
\citep{kokonendji2004a}, Poisson-Tweedie \citep{kokonendji2004b}, discrete
Lindley distribution \citep{gomez2011a}, a new logarithmic distribution 
\citep{gomez2011b}. discrete Gamma distribution \citep{chakraborty2012}, and
discrete generalized Rayleigh distribution \citep{alamatsaz2016}. See 
\citet{coly2016}, for a review on other models and applications.

Concerning the development of models to handle zero-inflation, we mention
zero-inflated Poisson \citep{lambert1952}, zero-inflated negative binomial 
\citep{ridout2001}, and a modeling based on copula functions 
\citep{zhao2012}, and for a review on more models and applications we refer to 
\citet{yip2005}. A few studies considered the development of models for both
overdispersed and zero-inflated data, for instance, generalized Poisson 
\citep{consul1989}, and geometric discrete Pareto distribution 
\citep{bhati2019}, exponentiated discrete Lindley distribution 
\citep{elmorshedy2020}.

All above mentioned models deal with parametric distributions. The
parameters are estimated directly from the data by moment matching or
maximum likelihood estimation. Another statistical way of estimating the
parameters of these distributions is by considering the concept of
generalized linear models. For instance, Poisson regression is a generalized
linear model with Poisson distribution error structure and the natural
logarithm link function. We refer to \citet{fahrmeir2006} for an overview of
regression models dealing with both overdispersed and zero-inflated data.

The approach that we followed for developing our fitted models is based on
the concept of natural exponential families, and its generalization to
exponential dispersion models \citep{jorgensen1997,barlev2017}. These models
play an important role in probabilistic and statistical modeling. Many
well-known distributions and families of distributions are included in these
models. Moreover, they show nice mathematical properties which are
convenient for the practical fitting of data. We defer theoretical details
on natural exponential families, exponential dispersion models, and our
usage of these models to Section \ref{s:families}. In Section 
\ref{s:computing} we will elaborate on the computation of the probabilities and
show some descriptive properties. Section \ref{s:data} concerns fitting our
models to real data and shows that our models perform good or best in all
cases when compared to best models reported in literature.


\section{Families of Probability Distributions}\label{s:families} 
In this section we first summarize the techniques of
modeling probability distributions based on the concepts of natural
exponential families, exponential dispersion models, and mean
parameterization. For more details we refer to \citet{barlev2017,letac1990}.
Also, we present our specific classes of these models.

\subsection{Natural Exponential Families}\label{ss:nef} 
Natural exponential families of distributions form a classic
tool in statistical modeling and analysis, e.g., see \citet{barndorff1978}.
We give the definition here to introduce our notation. Let $\mu$ be a
positive Radon measure on the Borel sets of the real line $\mathbb{R}$, with
convex support, and define $\Theta\doteq\big\{\theta\in\mathbb{R}: \int
e^{\theta x}\mu(dx)<\infty\big\}$. Assume that $\Theta$ is nonempty and
open. According to the H\"{o}lder's inequality, $\Theta$ is an interval. We
define for $\theta\in \Theta$ the cumulant transform 
\begin{equation*}
\kappa(\theta)\doteq \log \int e^{\theta x}\,\mu(dx).
\end{equation*}
Then, the natural exponential family $\mathcal{F}$, generated by $\mu$ is
defined by the set of probability distributions 
\begin{equation}  \label{e:nef}
\mathcal{F}\doteq \big\{ F_\theta: F_\theta(dx)\doteq e^{\theta
x-\kappa(\theta)}\,\mu(dx) : \theta \in \Theta\big\}.
\end{equation}
The measure $\mu$ is called the kernel of the family. The family is
parameterized by the natural parameter $\theta$. Differentiation is
permitted and gives the mean $m(\theta)=\kappa^{\prime}(\theta)$ and
variance $V(\theta)=\kappa^{\prime\prime}(\theta)$.

\subsection{Exponential Dispersion Models}\label{ss:edm} 
The concept of natural exponential families generalizes in
the following manner to exponential dispersion models which gives the
statistical practitioner a convenient framework \citep{jorgensen1997}. Given
a natural exponential family $\mathcal{F}$ with kernel $\mu$ and cumulant 
$\kappa(\theta)$, let $\Lambda$ be the set of $p>0$ such that 
$p\kappa(\theta) $ is the cumulant transform for some measure $\mu_p$. Thus
for any $p\in \Lambda$, we get a natural exponential family of the form 
\begin{equation*}
\mathcal{F}(\mu_p)\doteq \big\{ F_{(\theta,p)}: F_{(\theta,p)}(dx)\doteq
e^{\theta x-p\kappa(\theta)}\,\mu_p(dx) : \theta \in \Theta\big\}.
\end{equation*}
This family of distributions, parameterized by pairs $(\theta,p)
\in\Theta\times\Lambda$ is called the exponential dispersion model. 
Parameter $p$ is
called the dispersion parameter. Many useful families of discrete
distributions belong to exponential dispersion models 
\citep{kokonendji2004b}, and for this reason these models are are 
important for fitting discrete
data. Furthermore, exponential dispersion models show convenient properties,
such as infinite divisibility (iff $\lambda=(0,\infty)$), see
\citet{jorgensen1997} for more details.

\subsection{Mean Parameterization}\label{ss:mean} 
For our purposes it is convenient to consider a
reparameterization of the natural exponential families, 
as introduced in \citet{letac1990}. 
The image of the differentiation of the cumulant, i.e., 
$\mathcal{M}\doteq\kappa^{\prime}(\Theta)$, is called the mean domain of the
family $\mathcal{F}$. Because the map 
$\theta \mapsto \kappa^{\prime}(\theta)$ 
is one-to-one, its inverse function $\psi : \mathcal{M}\to \Theta$ is
well defined, i.e., 
\begin{equation}  \label{e:psi}
\psi(m)\doteq \big(\kappa^{\prime}\big)^{-1}(m),\; m\in \mathcal{M}.
\end{equation}
The corresponding variance function is 
$V(m)\doteq V\big(\psi(m)\big)$. 
Furthermore, we define 
\begin{equation}  \label{e:psi1}
\psi_1(m)\doteq \kappa\big(\psi(m)\big),\; m\in \mathcal{M}.
\end{equation}
In this way the natural exponential family $\mathcal{F}$ is modeled by its
mean parameterization, 
\begin{equation}  \label{e:nefm}
\mathcal{F}\doteq \big\{ F_m: F_m(dx)\doteq 
e^{\psi(m) x-\psi_1(m)}\,\mu(dx): m \in \mathcal{M} \big\}.
\end{equation}
Note that the mean parameterization has an associated variance 
function $V(m) $ on the mean domain $\mathcal{M}$. Conversely, when a natural
exponential family $\mathcal{F}$ is given just by a pair 
$\big(\mathcal{M},V(m)\big)$, we find the distributions by 
\begin{equation}  \label{e:primitives}
\psi(m)=\int\frac{1}{V(m)}\,dm;\quad \psi_1(m)=\int\frac{m}{V(m)}\,dm.
\end{equation}

\bigskip\noindent 
In our study we are interested in discrete distributions,
say with probability mass functions 
$f(n) = \mathbb{P}(X=n)$, $n=0,1,\ldots$, 
where $X$ represents the associated random variable. Specifically, we
consider distributions belonging to exponential dispersion models, using the
mean parameterization. Thus, the probability mass functions are represented
by 
\begin{equation}  \label{e:pmf}
f_m(n)\doteq \mu_ne^{n\psi(m)-\psi_1(m)},\; n=0,1,\ldots.
\end{equation}

\subsection{Variance Function Classes}\label{ss:our} 
Here we present the three models (or families) of
distributions that we introduced and analysed in \citet{barlev2020a}. The
models are defined through their variance function classes of the mean
parameterization.

\begin{itemize}
\item 
ABM class, named after \citet{awad2016}, with mean $m$, and variance
function 
\begin{equation}  \label{e:vfabm}
V(m) = m\Big(1+\frac{m}{p}\Big)^r,\, p>0, r=2,3,\ldots.
\end{equation}
Note that in case of $r=0$, we get the Poisson distribution, and $r=1$ gives
the negative binomial distribution. The case $r=2$ is called generalized
Poisson \citep{consul1989}, or Abel distribution \citep{letac1990}.

\item 
LMS class, named after \citet{letac1990}, with mean $m$ and variance
function 
\begin{equation}  \label{e:vflms}
V(m) = m\Big(1+\frac{m}{b}\Big)\Big(1+\frac{m}{p}\Big)^r,\, p>0,
r=1,2,\ldots.
\end{equation}
The case $r=1$ with $p=b/2$ is called also generalized negative binomial %
\citep{jain1971}.

\item 
LMNS class with mean $m$ and variance function 
\begin{equation}  \label{e:vflmns}
V(m) = \frac{m}{\Big(1-\frac{m}{p}\Big)^r}, \, p>m, r=1,2,\ldots.
\end{equation}
Note that the mean domain of this class is finite, $\mathcal{M}=(0,p)$. This
class refers also to \citet{letac1990}, however, contrarily to the previous
LMS class, the LMNS class belongs to a non-steep natural exponential family.
Steepness of a natural exponential family is a property of its cumulant
transform $\kappa(\theta)$ \citep{barndorff1978}.
\end{itemize}

\noindent 
It has been shown in \citet{letac1990} that the natural
exponential families associated with these classes of variance functions are
concentrated on the nonnegative integers. To our best knowledge these
distributional models have not been used before in statistical fitting to
real data, except for the trivial cases. We have considered in 
\citet{barlev2019} the Abel distribution for modeling insurance claim data.

\medskip\noindent 
It is easy to see that all these three classes show
overdispersion, $V(m)>m$, and that the dispersion increases with the 
power $r $ in the expressions \eqref{e:vfabm}-\eqref{e:vflmns} of the variance
functions. Moreover, in \citet{barlev2020a} we have shown that these three
classes satisfy also an increasing zero-inflation property. This means, when
we would denote the probability mass in zero by $P_r(0;p,m)$ given mean $m$,
dispersion parameter $p$ and power $r$ in the variance function, then 
\begin{equation*}
P_r(0;p,m) < P_{r+1}(0;p,m),\; r=1,2,\ldots.
\end{equation*}
These properties make the three classes a flexible tool for fitting
overdispersed, zero-inflated data.

\subsection{Fitting Data}\label{ss:fitting} 
Consider a data set of counts, $n_0,n_1,\ldots,n_K$,
meaning, $n_0$ observations of value 0, $n_1$ observations of value 1, etc.
Let $N=\sum_{k=0}^Kn_k$ be the total number of observations. The empirical
probability mass function is 
\begin{equation*}
p^{\mathrm{(emp)}}_k=\frac{n_k}{N},\;k=0,\ldots,K.
\end{equation*}
The empirical mean $\overline{x}$, variance $s^2$, $3$-rd central moment 
$m_3 $, skewness $b_1$, and index of dispersion $D$ are 
\begin{align*}
\overline{x} &=\frac1N\sum_{k=1}^K kn_k
=\sum_{k=1}^K k\,p^{\mathrm{(emp)}}_k,\\
s^2 &=\frac{1}{N-1}\sum_{k=0}^K (k-\overline{x})^2\,n_k 
=\frac{N}{N-1}\sum_{k=0}^K (k-\overline{x})^2\,p^{\mathrm{(emp)}}_k, \\
m_3 &= \frac1N\sum_{k=0}^K (k-\overline{x})^3\,n_k 
=\sum_{k=0}^K (k-\overline{x})^3\,p^{\mathrm{(emp)}}_k, \\
b_1 &=\frac{m_3}{s^3} \\
D &= \frac{s^2}{\overline{x}}.
\end{align*}
Furthermore, consider a (theoretical) probability model of a discrete random
variable $X$ on $\{0,1,\ldots\}$ with probability mass function 
\begin{equation*}
p^{\mathrm{(mod)}}_k\doteq\mathbb{P}(X=k),\, k=0,1,\ldots.
\end{equation*}
The objective is to find a `good fitting' model of the data. The performance
of a model is expressed through the following measures.

\begin{itemize}
\item 
The logarithm of the likelihood: 
\begin{equation*}
L = \log\prod_{k=0}^Kn_k p^{\mathrm{(mod)}}_k.
\end{equation*}

\item 
$\chi^2$ value; taken into account sufficient expected number in the
categories.

\item 
$p$-value of the $\chi^2$ quantile; taken into account the number of
parameters that are estimated from the data.

\item 
Root mean squared error (RMSE): 
\begin{equation*}
\sqrt{\frac{1}{K+1}\sum_{k=0}^K \big(n_k - Np_k^{\mathrm{mod}}\big)^2}.
\end{equation*}
\end{itemize}

\noindent
In the literature there are usages of a few other measures, such as mean
square error (MSE), Akaike information criterion (AIC), or Bayesian
information criterion (BIC). These are related to the measures mentioned
earlier, and do not give more information. Other measures would be for
instance, total variation, $l_2$-norm, or Kullback-Leibler divergence. We do
not take these into account because these are not common in the literature
that we considered for comparisons.

\medskip\noindent 
We will study fitting models obtained by exponential
dispersion models, given by the three variance function classes of 
Section \ref{ss:our}, and compare these with models proposed in literature. The
model parameters $m,p,b$ are estimated by maximum likelihood method.


\section{Computing Distributions}\label{s:computing} 
In this section we elaborate the numerical methods
concerning computation of our distributions. The general procedure is
described in \citet{letac1990} which goes as follows. Let be given a
variance function $V(m)$ of a natural exponential family of discrete
distributions, and recall the integral equations \eqref{e:primitives}.

\begin{enumerate}
\item 
We solve for primitives $\widetilde{\psi}$ and $\widetilde{\psi}_1$ of
the integral equations with zero constant of integration. All solutions are 
\begin{equation*}
\psi(m)=\widetilde{\psi}(m)+c_0;\; 
\psi_1(m)=\widetilde{\psi}_1(m)+d_0,\;\; c_0,d_0\in\mathbb{R}.
\end{equation*}
The specific constants $c_0$ and $d_0$ are obtained by the boundary
conditions 
\begin{equation}  \label{e:boundary}
\psi_1(0)=0 \quad\text{and}\quad \lim_{m\to 0} m\,e^{-\psi(m)}=1.
\end{equation}
Define 
\begin{equation}  \label{e:G}
G(m) = m\,e^{-\psi(m)}.
\end{equation}
The kernel $(\mu_n)_{n=0}^\infty$ is computed according to 
\begin{equation}  \label{e:kernel}
\begin{split}
\mu_0 &= e^{\psi_1(0)} \\
\mu_n &= \frac{1}{n!} \big(\frac{d}{dm}\big)^{n-1}
e^{\psi_1(m)}\psi_1^{\prime}(m)\big(G(m)\big)^n\Big|_{m=0},\quad
n=1,2,\ldots.
\end{split}
\end{equation}

\item 
Note that the $d_0$ constant of the $\psi_1$ function equals 
\begin{equation*}
d_0 = -\widetilde{\psi}_1(0).
\end{equation*}
It is convenient to introduce $\widetilde{\psi}_0$ and $\psi_0$ functions,
by 
\begin{equation}  \label{e:psi0}
\widetilde{\psi}_0(m)\doteq \widetilde{\psi}(m)-\log(m);\quad
\psi_0(m)\doteq \widetilde{\psi}_0(m)+c_0=\psi(m)-\log(m).
\end{equation}
Then the $c_0$ constant of the $\psi$ function equals 
\begin{equation*}
c_0 = -\widetilde{\psi}_0(0),
\end{equation*}
and $G(m)=e^{-\psi_0(m)}$.

\item 
The main difficulty lies in computing the kernel as in \eqref{e:kernel}. 
Define for $n=1,2,\ldots$ 
\begin{equation*}
F_n(m) \doteq e^{\psi_1(m)}\,\psi_1^{\prime}(m)\,\big(G(m)\big)^n,
\end{equation*}
and $H_n(m) \doteq \log F_n(m)$. Thus, 
\begin{equation*}
H_n(m) = \psi_1(m) + \log\psi^{\prime}(m) - n\psi_0(m).
\end{equation*}
It is easy to see that 
\begin{equation*}
H_n(0)=\psi_1(0) + \log\psi_1^{\prime}(0) -n\psi_0(0) =0 + 
\log 1 - n (\widetilde{\psi}_0(0)+c_0) = 0.
\end{equation*}
The next step is to compute the $(n-1)$-st derivative of $F_n(m)$ 
in \eqref{e:kernel}. Apply the chain rule: 
\begin{align*}
&\big(\frac{d}{dm}\big)^{n-1} F_n(m) = 
\big(\frac{d}{dm}\big)^{n-1}e^{H_n(m)} \\
&= \sum \frac{(n-1)!}{k_1!k_2!\cdots k_{n-1}!}\, 
\Big(\big(\frac{d}{dx}\big)^k e^x\Big)_{x=H_n(m)}\, 
\prod_{j=1}^{n-1}\Big(\frac{H_n^{(j)}(m)}{j!}\Big)^{k_j},
\end{align*}
where the sum is over all nonnegative integer solutions of the Diophantine
equation $\sum_{j=1}^{n-1}jk_j=n-1$, and where $k=\sum_{j=1}^{n-1}k_j$. In
the next sections we compute the derivatives $H_n^{(j)}(m)$ for the separate
classes. The final computation for kernel $\mu_n$ involves evaluating these
expressions in $m=0$, and dividing by $n!$. Thus, 
\begin{align}
& \mu_n = \frac{1}{n!}\Big(\big(\frac{d}{dm}\big)^{n-1}
e^{\psi_1(m)}\,\psi_1^{\prime}(m)\,\big(G(m)\big)^n\Big)\Big|_{m=0}  
\notag \\
&=\frac{1}{n!}\Big(\big(\frac{d}{dm}\big)^{n-1} e^{H_n(m)}\Big)\Big|_{m=0} 
\notag \\
&= \frac{1}{n!} \sum \frac{(n-1)!}{k_1!k_2!\cdots k_{n-1}!}\, 
\Big(\big(\frac{d}{dx}\big)^k e^x\Big)_{x=H_n(0)}\, 
\prod_{j=1}^{n-1} \Big(\big(\frac{H_n^{(j)}(m)}{j!}\big)\big|_{m=0}\Big)^{k_j}  
\notag \\
&=\frac1n \sum \frac{1}{k_1!k_2!\cdots k_{n-1}!}\, 
\prod_{j=1}^{n-1}\Big(\frac{H_n^{(j)}(0)}{j!}\Big)^{k_j}  
\notag \\
&=\frac1n \sum \prod_{j=1}^{n-1}\prod_{t=1}^{k_j} \frac{H_n^{(j)}(0)/j!}{t}.
\label{e:mun}
\end{align}
\end{enumerate}

\subsection{Computing ABM Distribution}\label{ss:abm} 
Recall that the variance function is given by 
\begin{equation*}
V(m)=m\big(1+\frac{m}{p}\big)^{r}, \, r=2,3,\ldots.
\end{equation*}
The integral equations are solved by \citep{barlev2020a} 
\begin{align*}
\widetilde{\psi}(m)&= \log(m) + \sum_{i=1}^{r-1}c_i\,(m+p)^{-i} +
c_r\log(m+p); \\
\widetilde{\psi}_1(m)&= d_{r-1}\,(m+p)^{-r+1},
\end{align*}
where the coefficients $c_i,d_{r-1}$ are, 
\begin{align*}
c_i &= 
\begin{cases}
\frac{p^i}{i} & \quad i=1,\ldots,r-1; \\ 
-1 & \quad i=r;
\end{cases}
\\
d_{r-1}&=-p^r/(r-1).
\end{align*}
Hence, 
\begin{equation*}
\psi_1^{\prime}(m)=\widetilde{\psi}_1^{\prime}(m) =p^r (m+p)^{-r}=\exp\big(%
r\log(p) - r\log(m+p)\big).
\end{equation*}
We get 
\begin{align*}
& H_n(m) = \psi_1(m) + \log\psi_1^{\prime}(m) - n\psi_0(m) \\
&=d_{r-1}(m+p)^{-r+1}+d_0 + r\log(p) - r\log(m+p) -
n\sum_{i=1}^{r-1}c_i(m+p)^{-i} -nc_r\log(m+p)-nc_0 \\
&=q_0 + \sum_{i=1}^{r-1}q_i(m+p)^{-i} + q_r\log(m+p) ,
\end{align*}
with 
\begin{align*}
q_0 &= d_0 +r\log(p)-nc_0 \\
q_i &= -nc_i,\;\; i=1,\ldots,r-2 \\
q_{r-1} &= d_{r-1}-nc_{r-1} \\
q_r &= -r-nc_r
\end{align*}
The $j$-th derivative of the $H_n(m)$ function becomes 
\begin{equation*}
H_n^{(j)}(m) = \sum_{i=1}^{r-1}q_i \big(\frac{d}{dm}\big)^j (m+p)^{-i} +q_r%
\big(\frac{d}{dm}\big)^j \log(m+p).
\end{equation*}
Thus, by defining 
\begin{equation*}
h_n(i,j)= 
\begin{cases}
\frac{1}{j!}\Big(\big(\frac{d}{dm}\big)^j (m+p)^{-i}\Big)\Big|_{m=0}
=(-1)^j\,\binom{j+i-1}{j}\,p^{-i-j} & \quad i=1,\ldots,r-1; \\ 
\frac{1}{j!}\Big(\big(\frac{d}{dm}\big)^j \log(m+p)\Big)\Big|_{m=0}
=(-1)^{j-1}\,\frac{1}{j}\,p^{-j} & \quad i=r,
\end{cases}
\end{equation*}
we obtain for the kernel $\mu_n$ in \eqref{e:mun}, 
\begin{equation*}
\mu_n =\frac1n \sum \prod_{j=1}^{n-1}\prod_{t=1}^{k_j}\sum_{i=1}^{r} \frac{%
q_ih_n(i,j)}{t}.
\end{equation*}

\subsection{Computing LMS Distribution}\label{ss:lms} 
The variance function is 
\begin{equation*}
V(m)=m\big(1+\frac{m}{b}\big)\big(1+\frac{m}{p}\big)^{r}, \, r=1,2,\ldots.
\end{equation*}
The integral solutions are \citep{barlev2020a} 
\begin{align*}
\widetilde{\psi}(m)&= \log(m) + \sum_{i=1}^{r-1}c_i\,(m+p)^{-i} +
c_r\log(m+p) + c_{r+1}\log(m+b); \\
\widetilde{\psi_1}(m)&= \sum_{i=1}^{r-1}d_i(m+p)^{-i} + d_r\log(m+p) +
d_{r+1}\log(m+b),
\end{align*}
where the coefficients $c_i,d_i$, $i=1,\ldots,r$ are, 
\begin{align*}
c_i &= 
\begin{cases}
\frac{p^i}{i}\Big(1 - \big(\frac{p}{p-b}\big)^{r-i} \Big) & \quad
i=1,\ldots,r-1; \\ 
\big(\frac{p}{p-b}\big)^r-1 & \quad i=r; \\ 
-\big(\frac{p}{p-b}\big)^r & \quad i=r+1
\end{cases}
\\
d_i &= 
\begin{cases}
b\frac{p^i}{i}\big(\frac{p}{p-b}\big)^{r-i} & \quad i=1,\ldots,r-1; \\ 
-b\Big(\frac{p}{p-b}\Big)^r & \quad i=r; \\ 
b\Big(\frac{p}{p-b}\Big)^r & \quad i=r+1.
\end{cases}
\end{align*}
Then, 
\begin{align*}
& \psi_1^{\prime}(m)=\widetilde{\psi}^{\prime}(m)=
-\sum_{i=1}^{r-1}id_i(m+p)^{-(i+1)} + d_r(m+p)^{-1} + d_{r+1}(m+b)^{-1} \\
&=\exp\big(\log(bp^r) - r\log(m+p) - \log(m+b)\big).
\end{align*}
This leads to 
\begin{align*}
& H_n(m) = \psi_1(m) + \log\psi_1^{\prime}(m) - n\psi_0(m) \\
&=q_0 + \sum_{i=1}^{r-1}q_i(m+p)^{-i} + q_r\log(m+p) +q_{r+1}\log(m+b),
\end{align*}
with 
\begin{align*}
q_0 &= d_0 +\log(bp^r)-nc_0 \\
q_i &= d_i -nc_i,\;\; i=1,\ldots,r-1 \\
q_{r} &= d_r -r -nc_r \\
q_{r+1} &= d_{r+1} -1-nc_{r+1}
\end{align*}
From here on, it goes similarly as in the computation of the ABM
distribution in Section \ref{ss:abm}. The kernel can be computed by 
\begin{equation*}
\mu_n=\frac1n \sum \prod_{j=1}^{n-1}\prod_{t=1}^{k_j}\sum_{i=1}^{r+1} 
\frac{q_ih_n(i,j)}{t},
\end{equation*}
where the first summation is over all nonnegative integer solutions of the
Diophantine equation $\sum_{j=1}^{n-1}jk_j=n-1$. The $h_n(i,j)$ constants
are defined by 
\begin{equation*}
h_n(i,j)= 
\begin{cases}
\frac{1}{j!}\Big(\big(\frac{d}{dm}\big)^j (m+p)^{-i}\Big)\Big|_{m=0}
=(-1)^j\,\binom{j+i-1}{j}\,p^{-i-j} & \quad i=1,\ldots,r-1; j=1,\ldots,n-1;
\\ 
\frac{1}{j!}\Big(\big(\frac{d}{dm}\big)^j \log(m+p)\Big)\Big|_{m=0}
=(-1)^{j-1}\,\frac{1}{j}\,p^{-j} & \quad i=r; j=1,\ldots,n-1; \\ 
\frac{1}{j!}\Big(\big(\frac{d}{dm}\big)^j \log(m+b)\Big)\Big|_{m=0}
=(-1)^{j-1}\,\frac{1}{j}\,b^{-j} & \quad i=r+1; j=1,\ldots,n-1.
\end{cases}
\end{equation*}

\subsection{Computing LMNS Distribution}\label{ss:lmns} 
The variance function is 
\begin{equation*}
V(m)=m\big(1-\frac{m}{p}\big)^{-r}, \, r=1,2,\ldots, 0<m<p.
\end{equation*}
The procedure is similar to computing the ABM and LMS distributions. Now it
holds that \citep{barlev2020a} 
\begin{align*}
\widetilde{\psi}(m)&= \log(m) + \sum_{i=1}^{r}c_i\,m^i; \\
\widetilde{\psi}_1(m)&= d_{r+1}(p-m)^{r+1},
\end{align*}
where 
\begin{align*}
c_i&= (-1)^i\binom{r}{i}\frac{1}{ip^i}\quad i=1,\ldots,r; \\
d_{r+1}&=-\frac{1}{(r+1)p^r}.
\end{align*}
Thus 
\begin{equation*}
\widetilde{\psi}_1^{\prime}(m)=\frac{1}{p^r}\,(p-m)^{r} =\exp\big(-r\log(p)
+ r\log(p-m)\big).
\end{equation*}
Now we get 
\begin{align*}
& H_n(m) = \psi_1(m) + \log\psi_1^{\prime}(m) - n\psi_0(m) \\
&= q_0 + \sum_{i=1}^rq_im^i + q_{r+1}(p-m)^{r+1}+q_{r+2}\log(p-m),
\end{align*}
with 
\begin{align*}
q_0 &= d_0 -r\log(p)-nc_0 \\
q_i &= -nc_i\;\; i=1,\ldots,r \\
q_{r+1} &= d_{r+1} \\
q_{r+2} &= r
\end{align*}
And, 
\begin{equation*}
\mu_n = \frac1n \sum \prod_{j=1}^{n-1}\prod_{t=1}^{k_j}\sum_{i=1}^{r+2} 
\frac{q_ih_n(i,j)}{t},
\end{equation*}
where 
\begin{equation*}
h_n(i,j)= 
\begin{cases}
\frac{1}{j!}\Big(\big(\frac{d}{dm}\big)^j m^i\Big)\Big|_{m=0} & \quad
i=1,\ldots,r; j=1,\ldots,n-1; \\ 
\frac{1}{j!}\Big(\big(\frac{d}{dm}\big)^j (p-m)^{r+1}\Big)\Big|_{m=0} & 
\quad i=r+1; j=1,\ldots,n-1; \\ 
\frac{1}{j!}\Big(\big(\frac{d}{dm}\big)^j \log(p-m)\Big)\Big|_{m=0} & \quad
i=r+2; j=1,\ldots,n-1.
\end{cases}
\end{equation*}
This becomes 
\begin{equation*}
h_n(i,j)= 
\begin{cases}
1 & \quad i=1,\ldots,r; j=i; \\ 
0 & \quad i=1,\ldots,r; j\not=i; \\ 
(-1)^j\binom{r+1}{j}p^{r+1-j} & \quad i=r+1; j=1,\ldots,\min\{r+1,n-1\}; \\ 
0 & \quad i=r+1; j>r+1; \\ 
-\frac{1}{jp^j} & \quad i=r+2; j=1,\ldots,n-1.
\end{cases}
\end{equation*}

\subsection{Histograms}\label{ss:hist} 
Before we discuss the results of the data analysis and the
fit of our distributions, we present a few histograms of these distributions
as an illustration of their shape and properties. From these figures we
observe that the probability mass functions are either decreasing or
unimodal. Furthermore, we see that all classes allow for zero-inflation, but
also for skewness to the right with less heavy probability mass in zero.
Generally, we conclude that our framework of the three variance function
classes for exponential families of distributions is a flexible tool for
describing discrete data.

\subsubsection{ABM}
\includegraphics[width=0.32\textwidth]{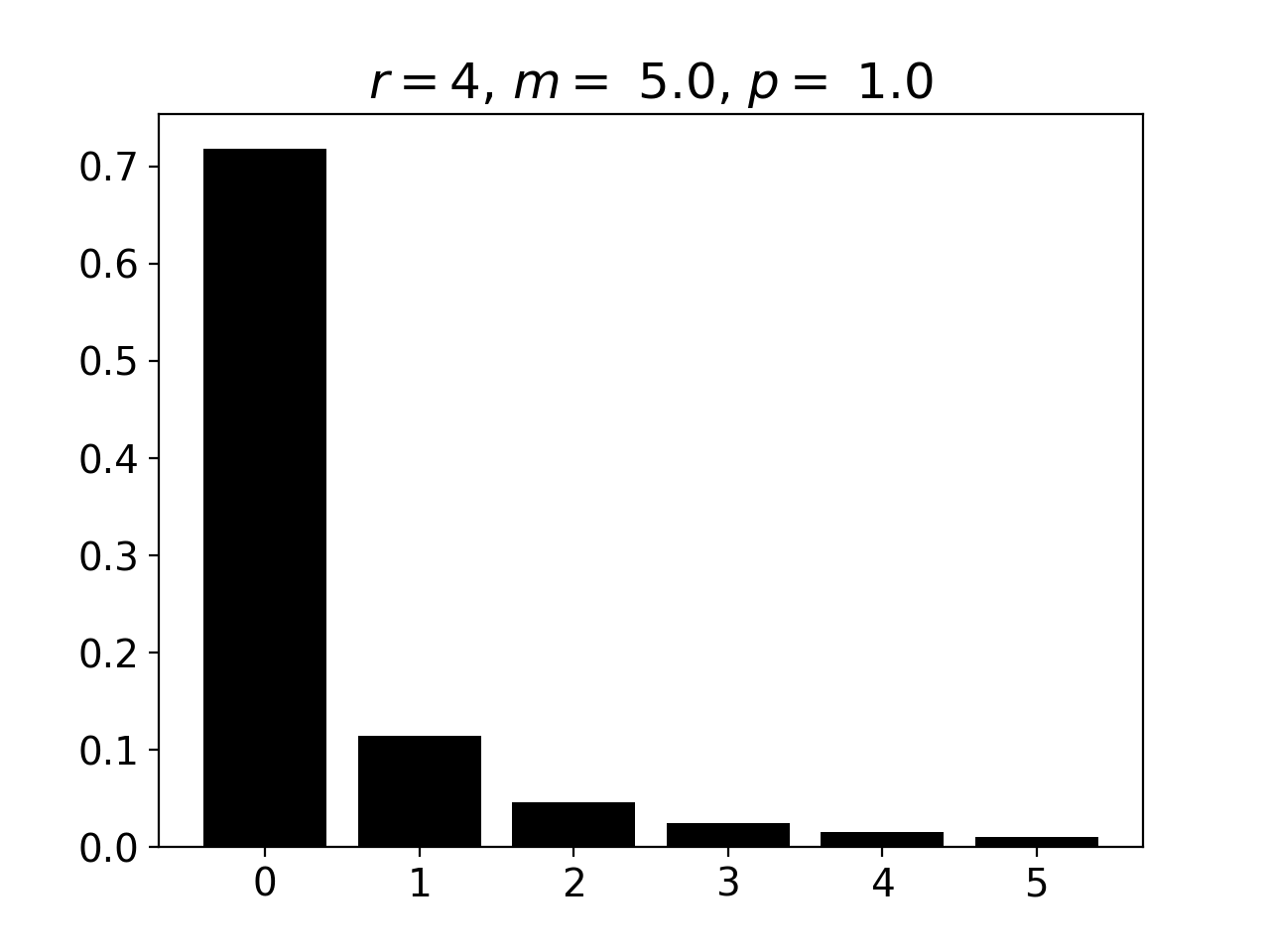}
\includegraphics[width=0.32\textwidth]{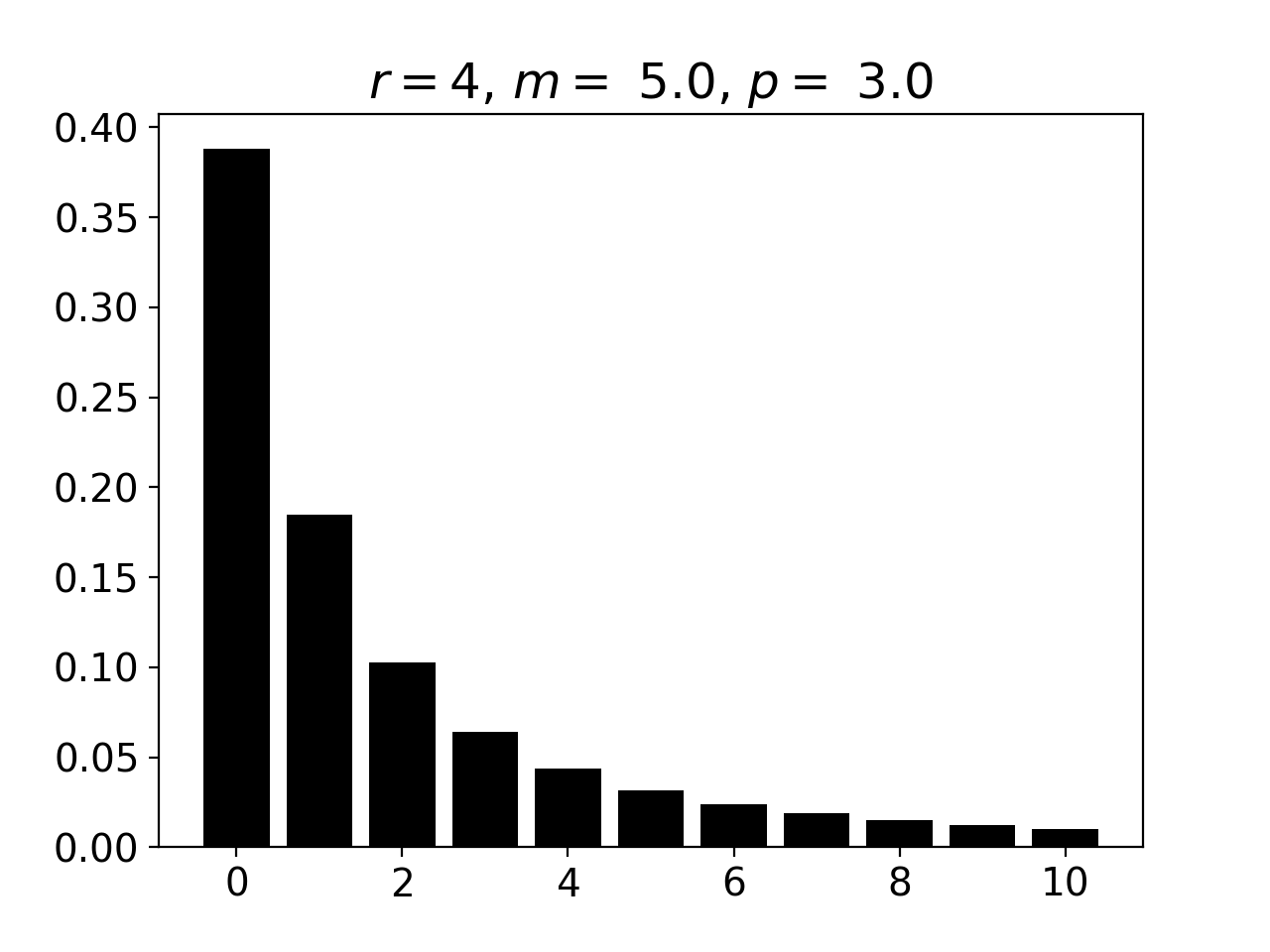}
\includegraphics[width=0.32\textwidth]{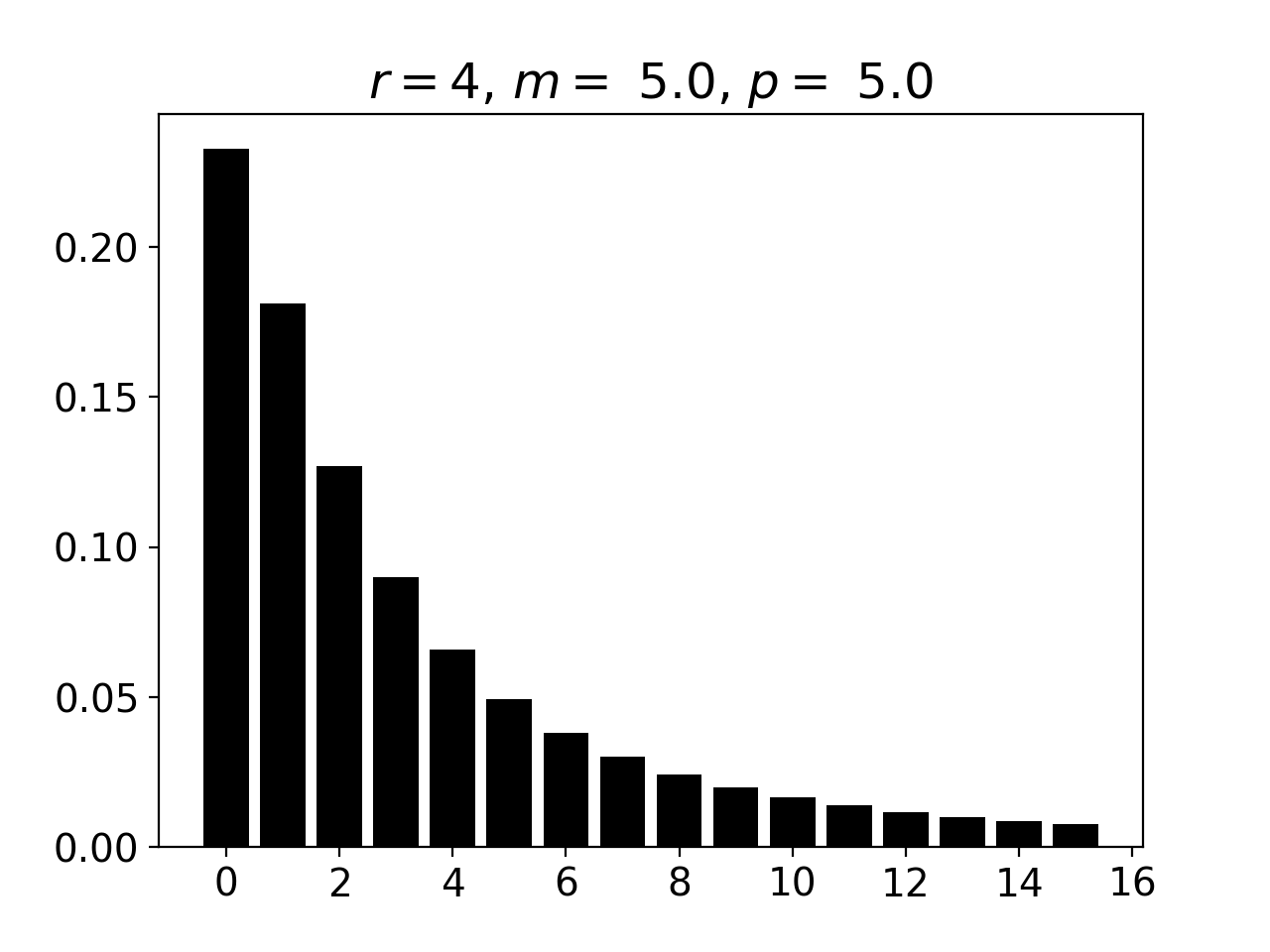}

\smallskip\noindent 
\includegraphics[width=0.32\textwidth]{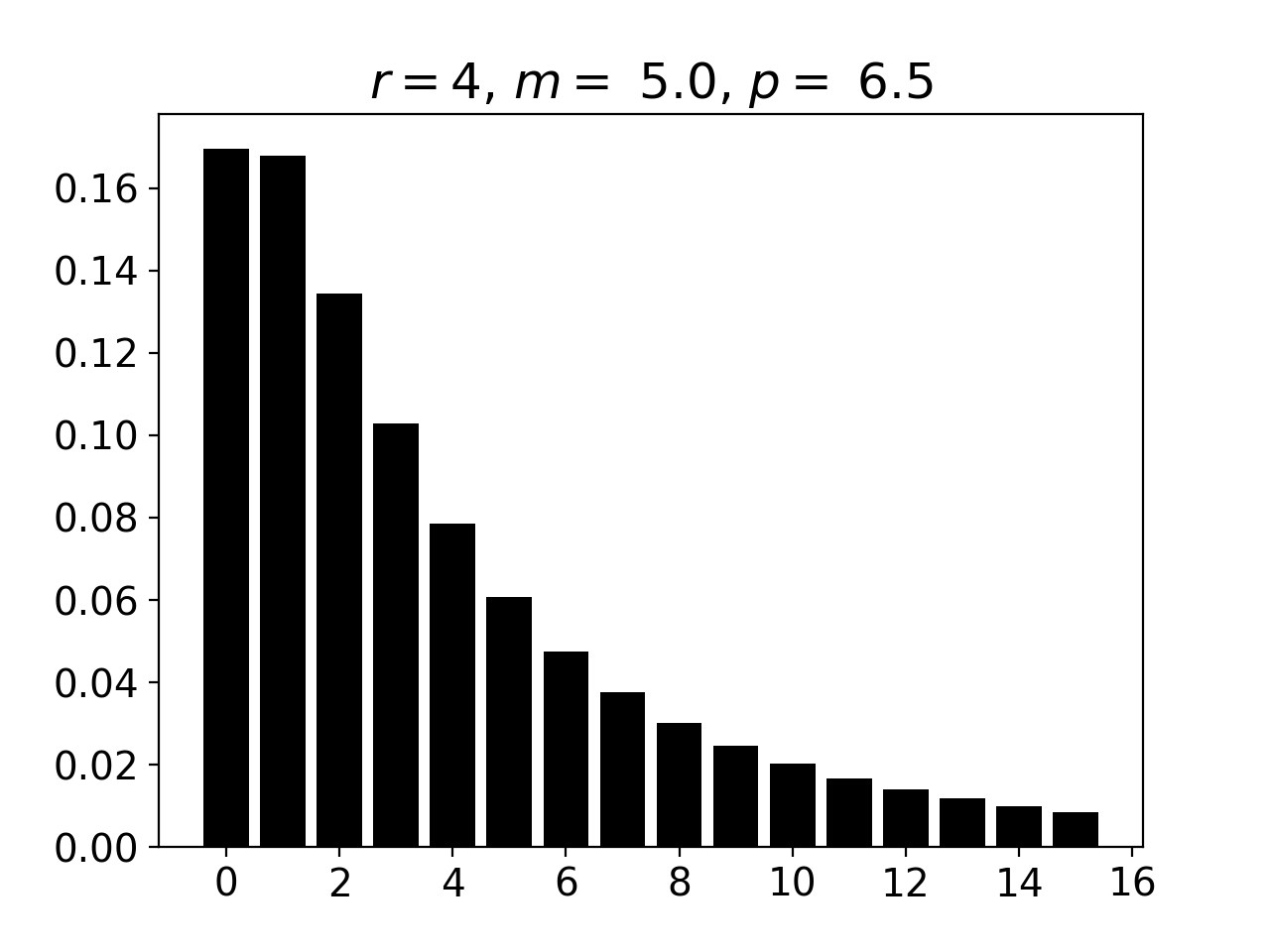}
\includegraphics[width=0.32\textwidth]{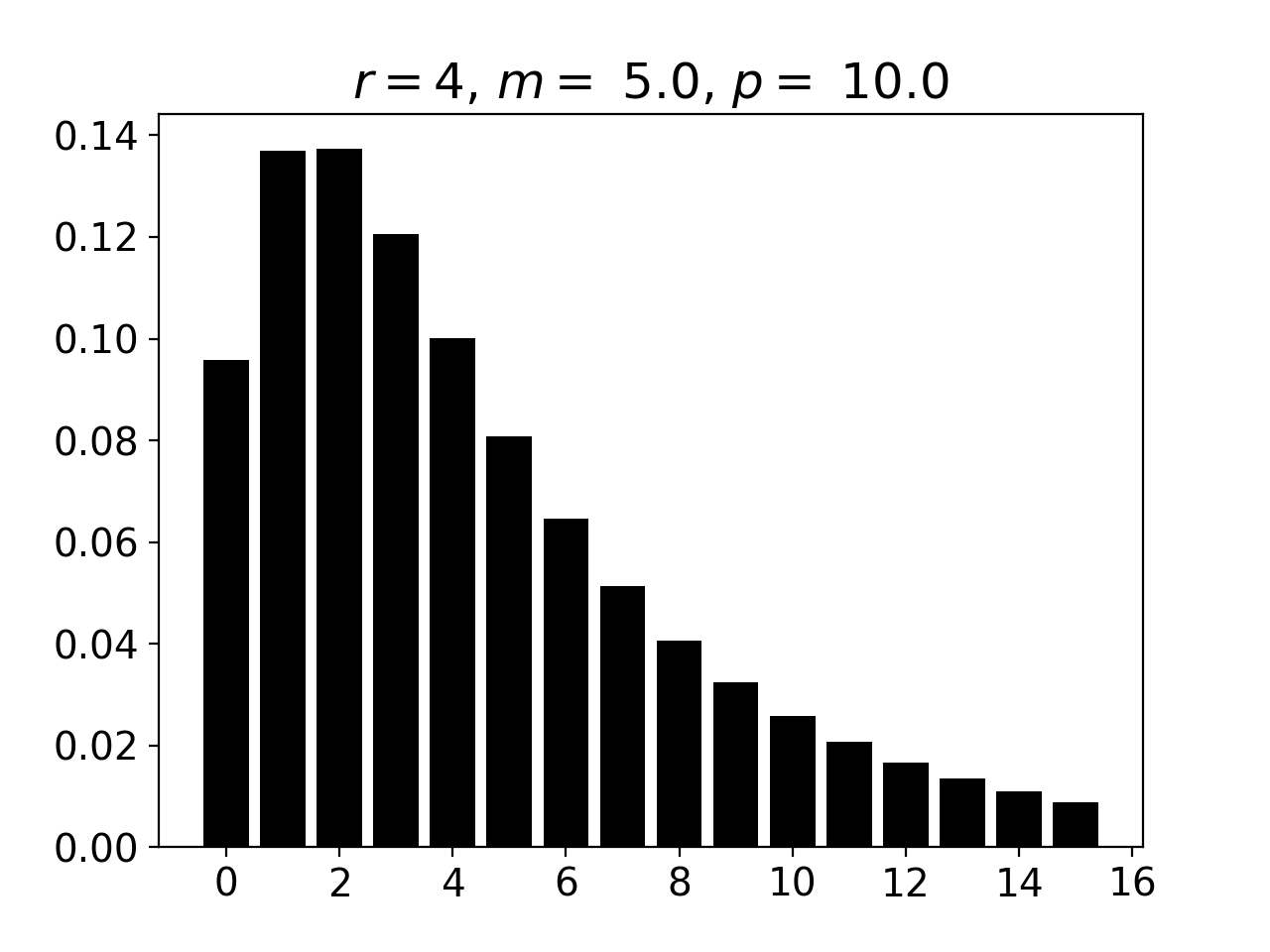}
\includegraphics[width=0.32\textwidth]{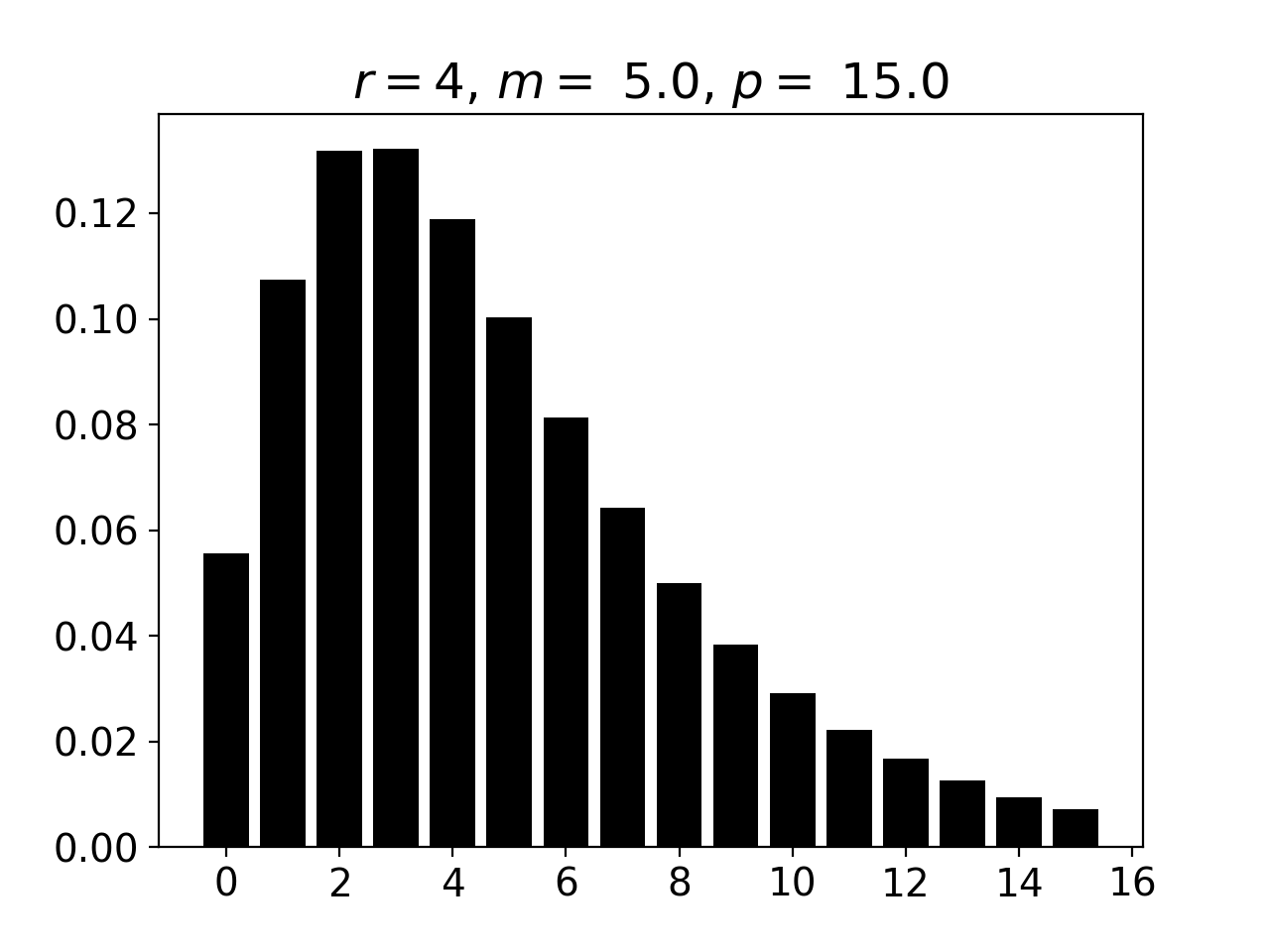}

\subsubsection{LMS}
\includegraphics[width=0.32\textwidth]{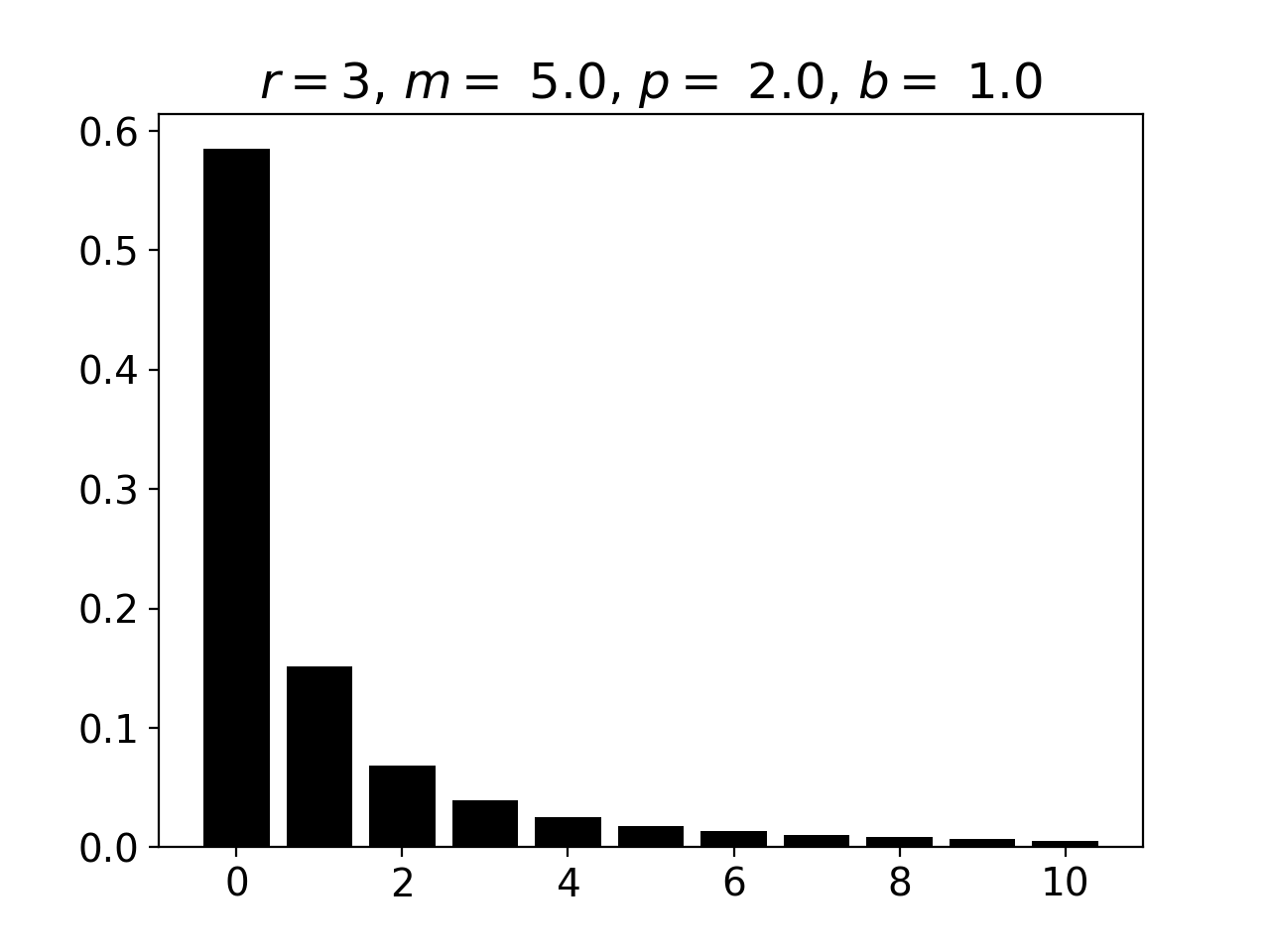}
\includegraphics[width=0.32\textwidth]{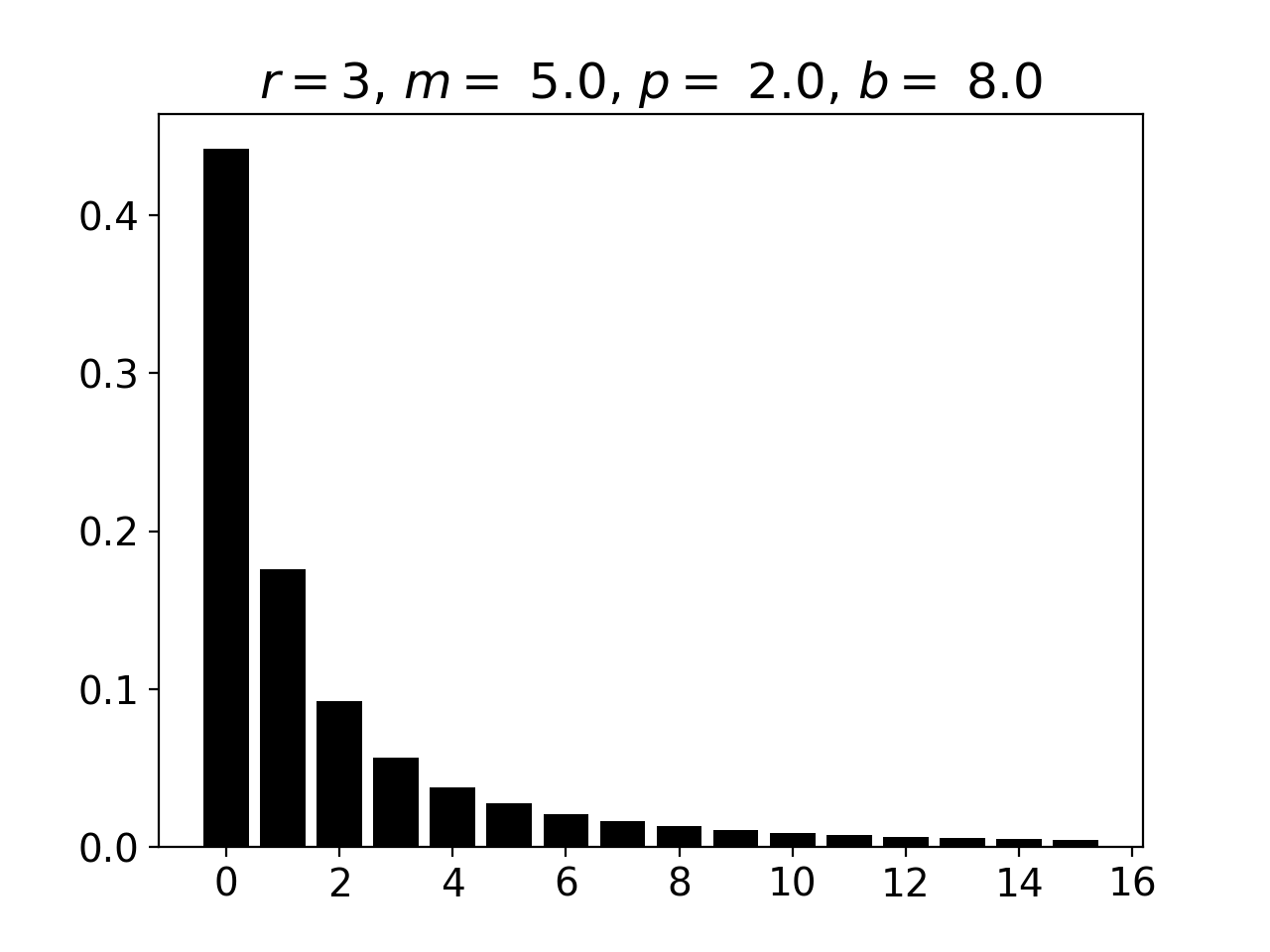}
\includegraphics[width=0.32\textwidth]{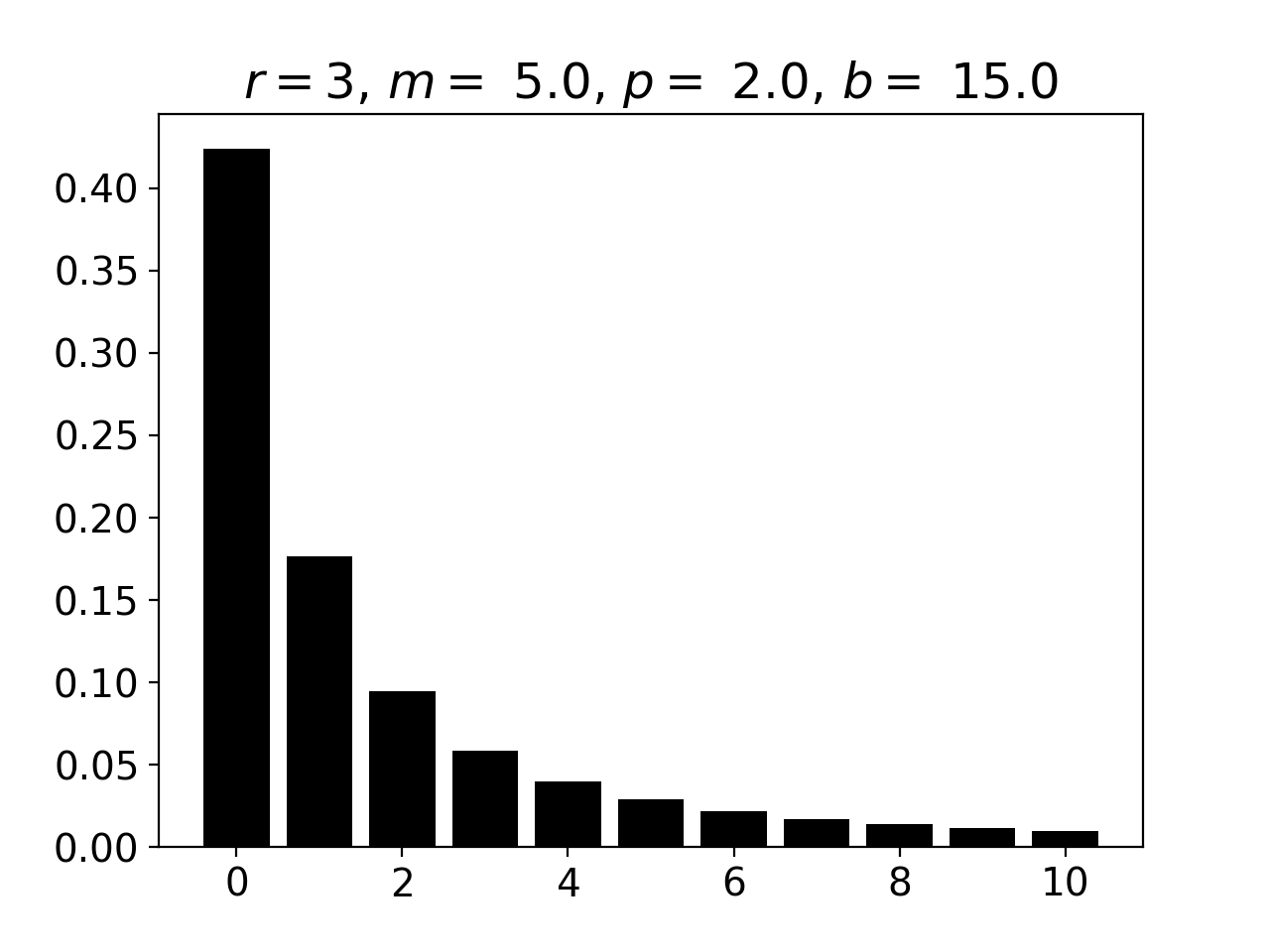}

\smallskip\noindent 
\includegraphics[width=0.32\textwidth]{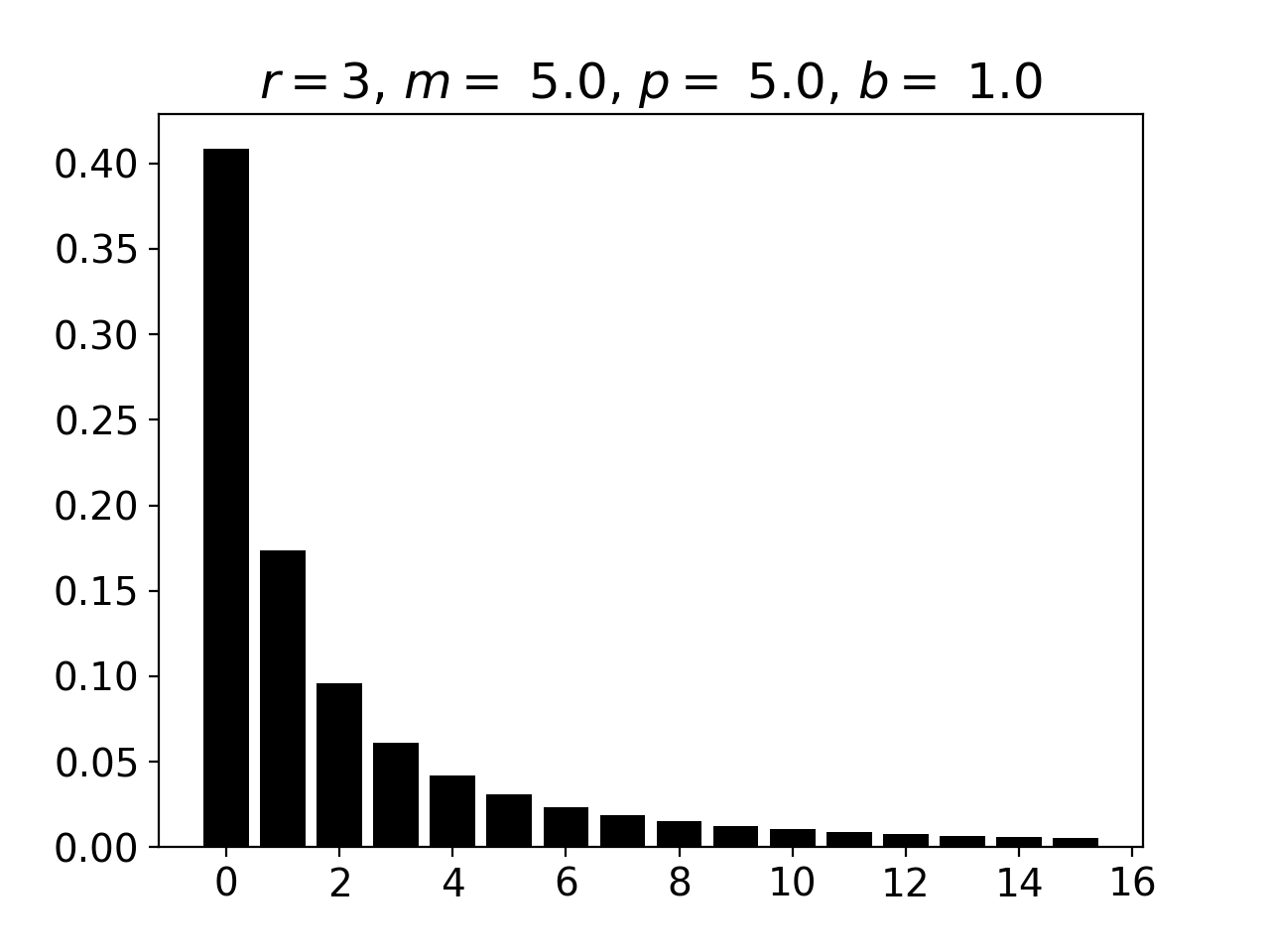}
\includegraphics[width=0.32\textwidth]{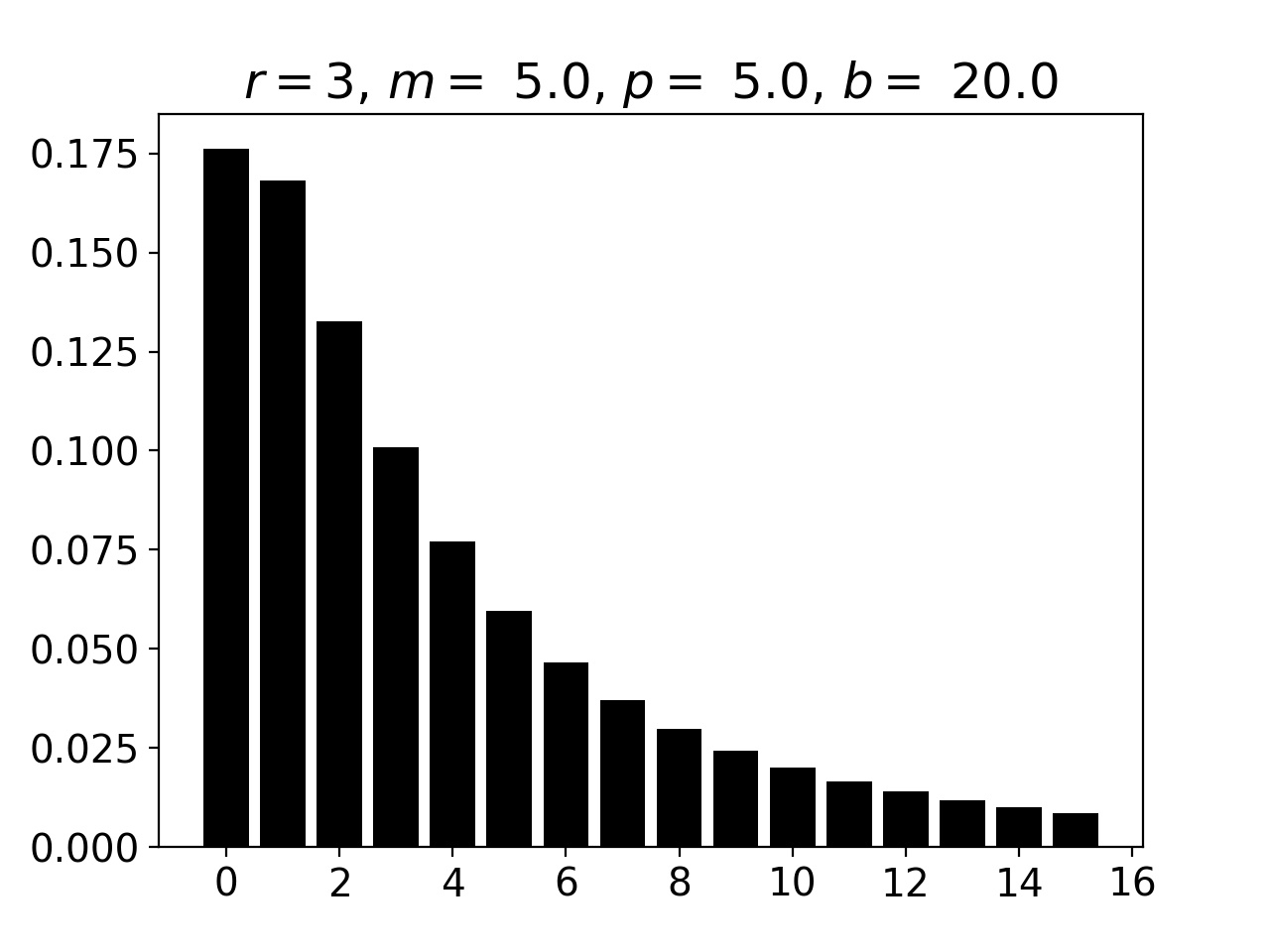}
\includegraphics[width=0.32\textwidth]{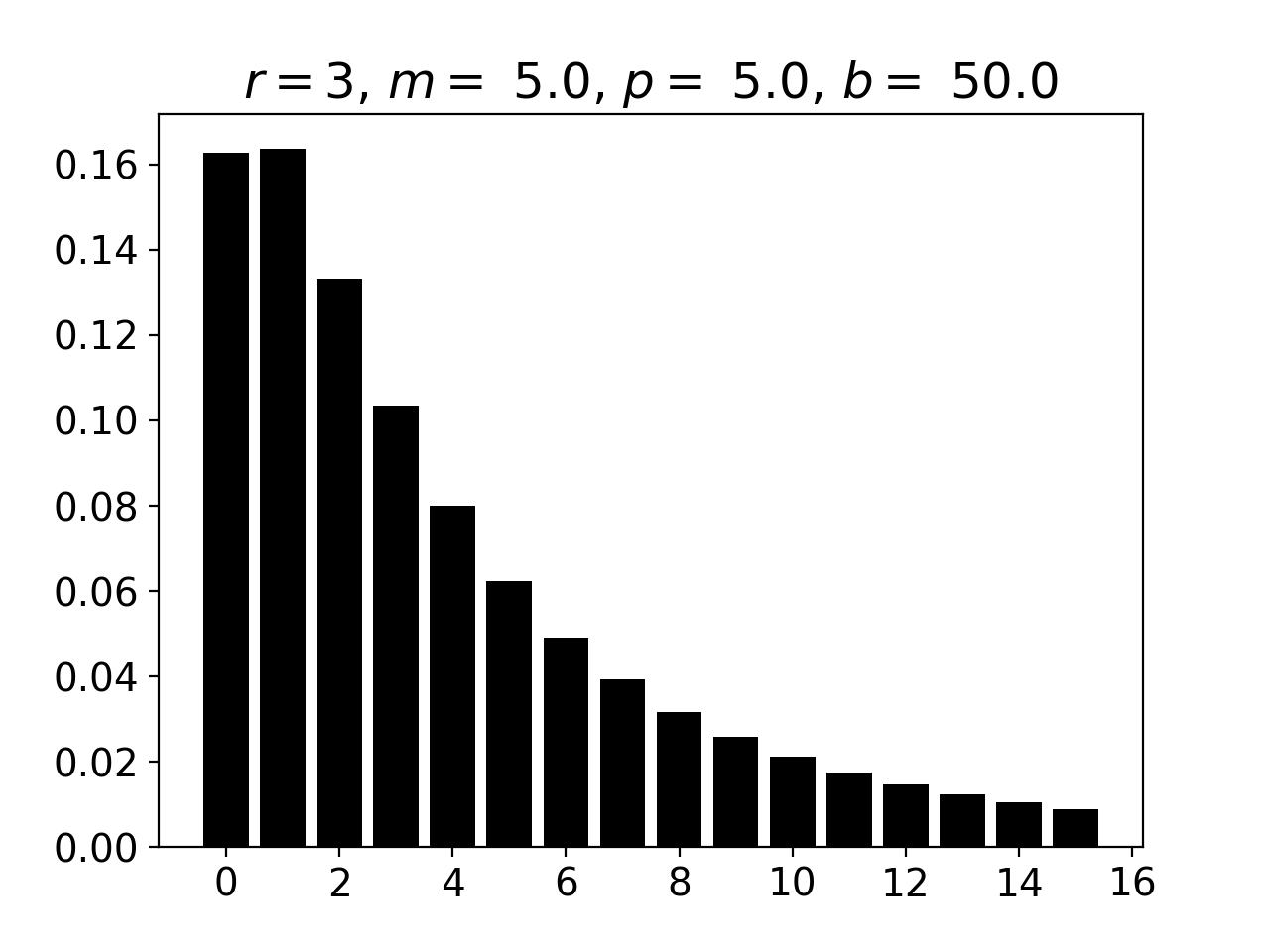}

\smallskip\noindent 
\includegraphics[width=0.32\textwidth]{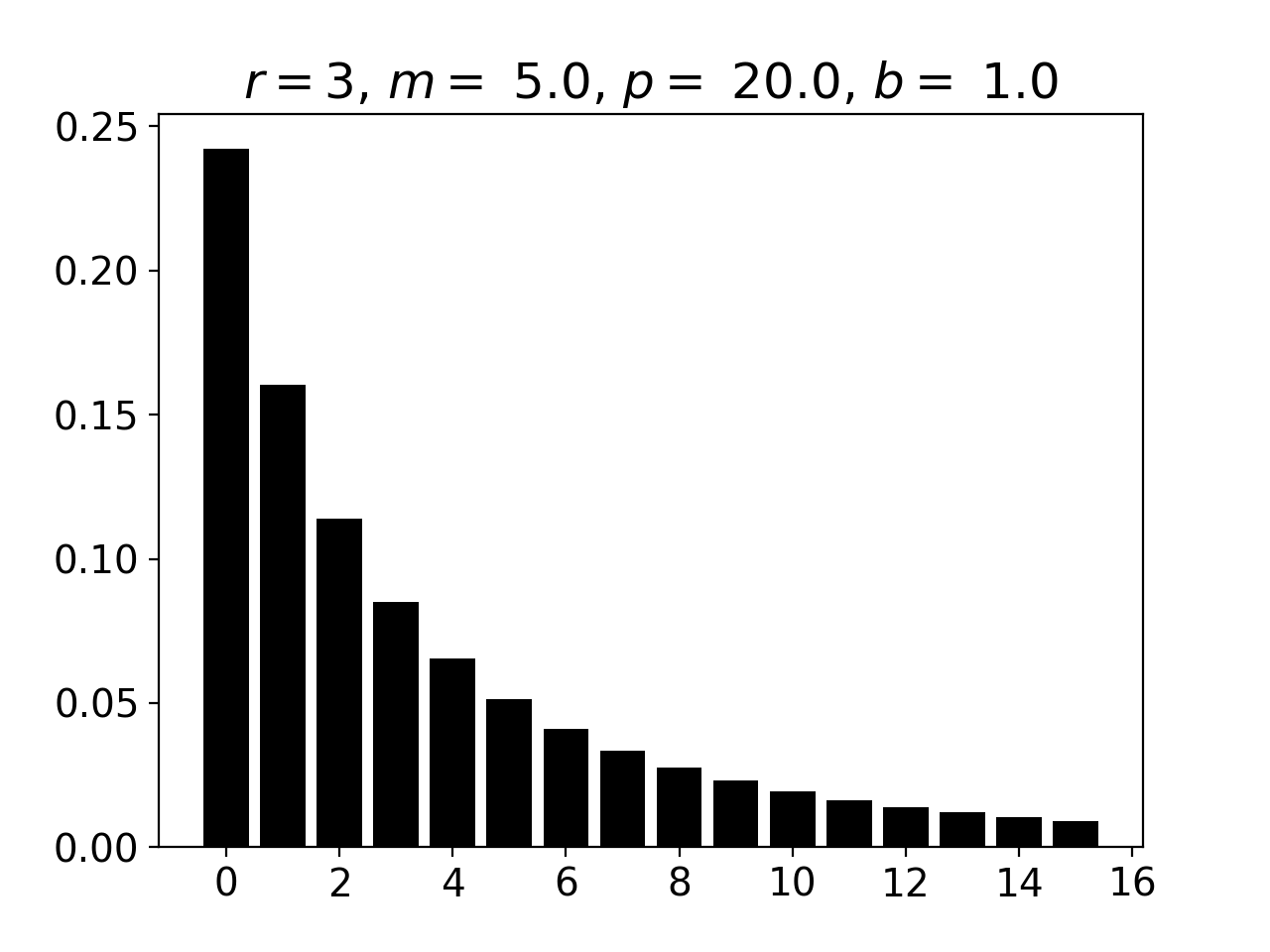}
\includegraphics[width=0.32\textwidth]{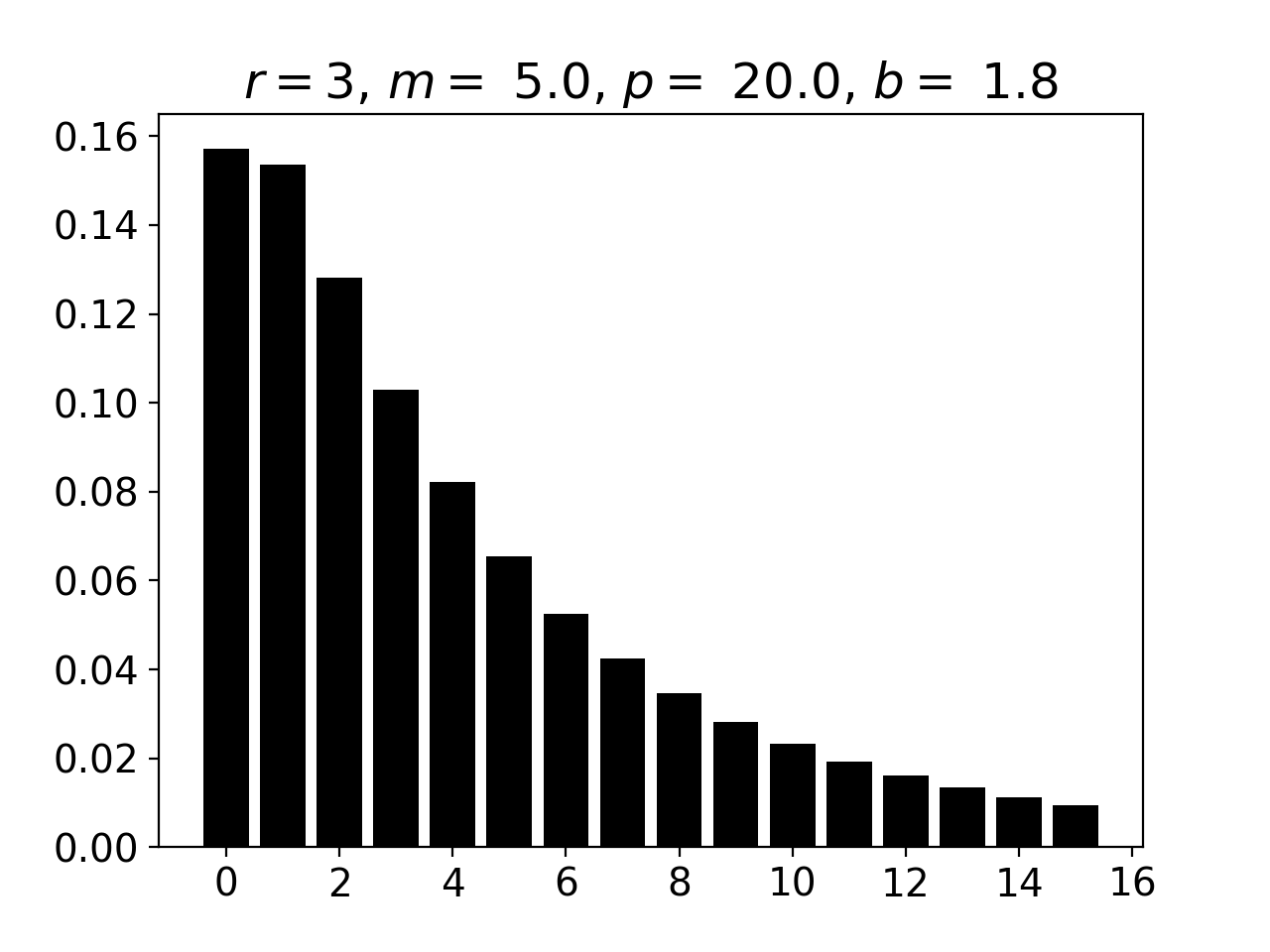}
\includegraphics[width=0.32\textwidth]{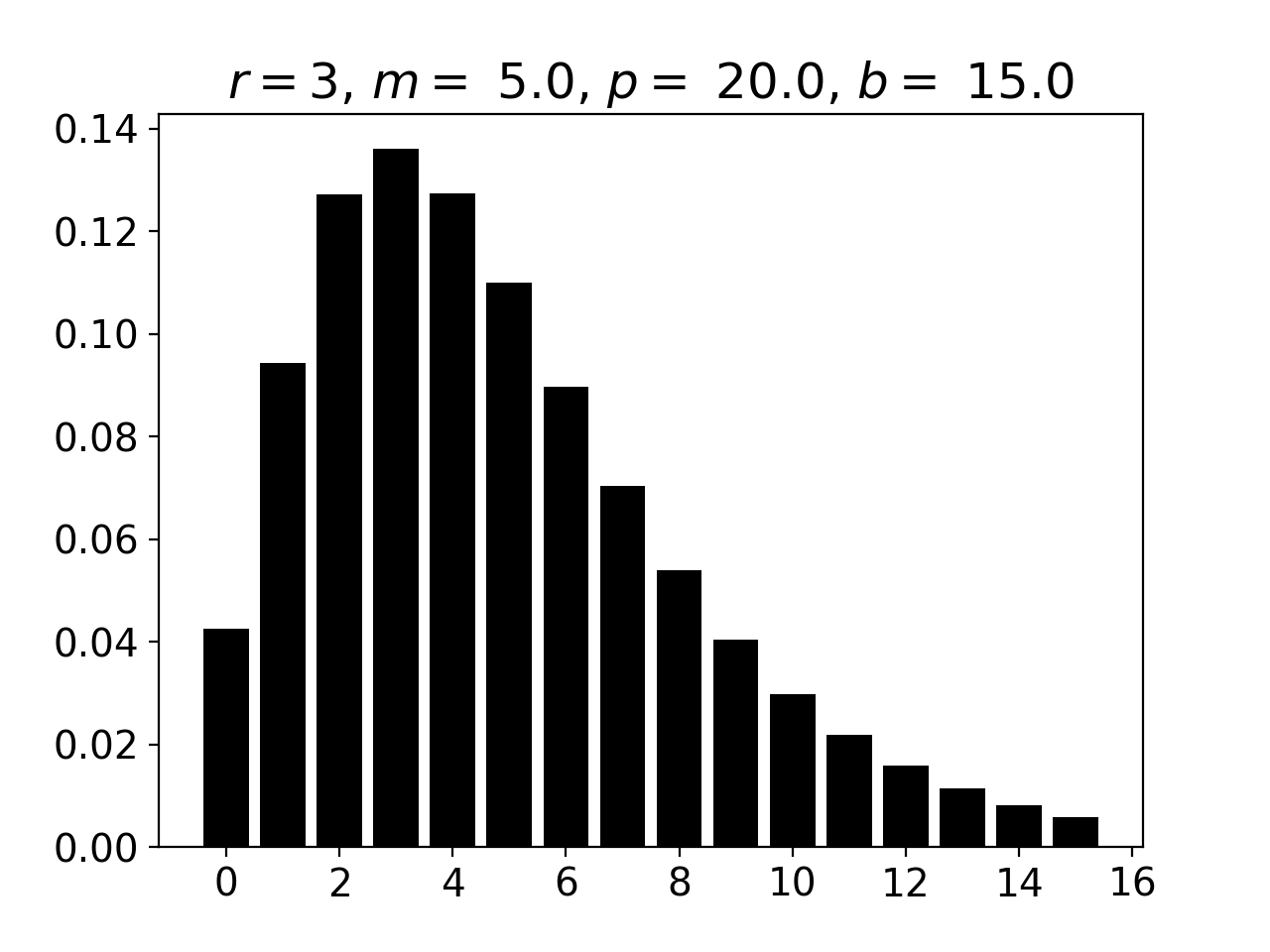}

\subsubsection{LMNS}
\includegraphics[width=0.32\textwidth]{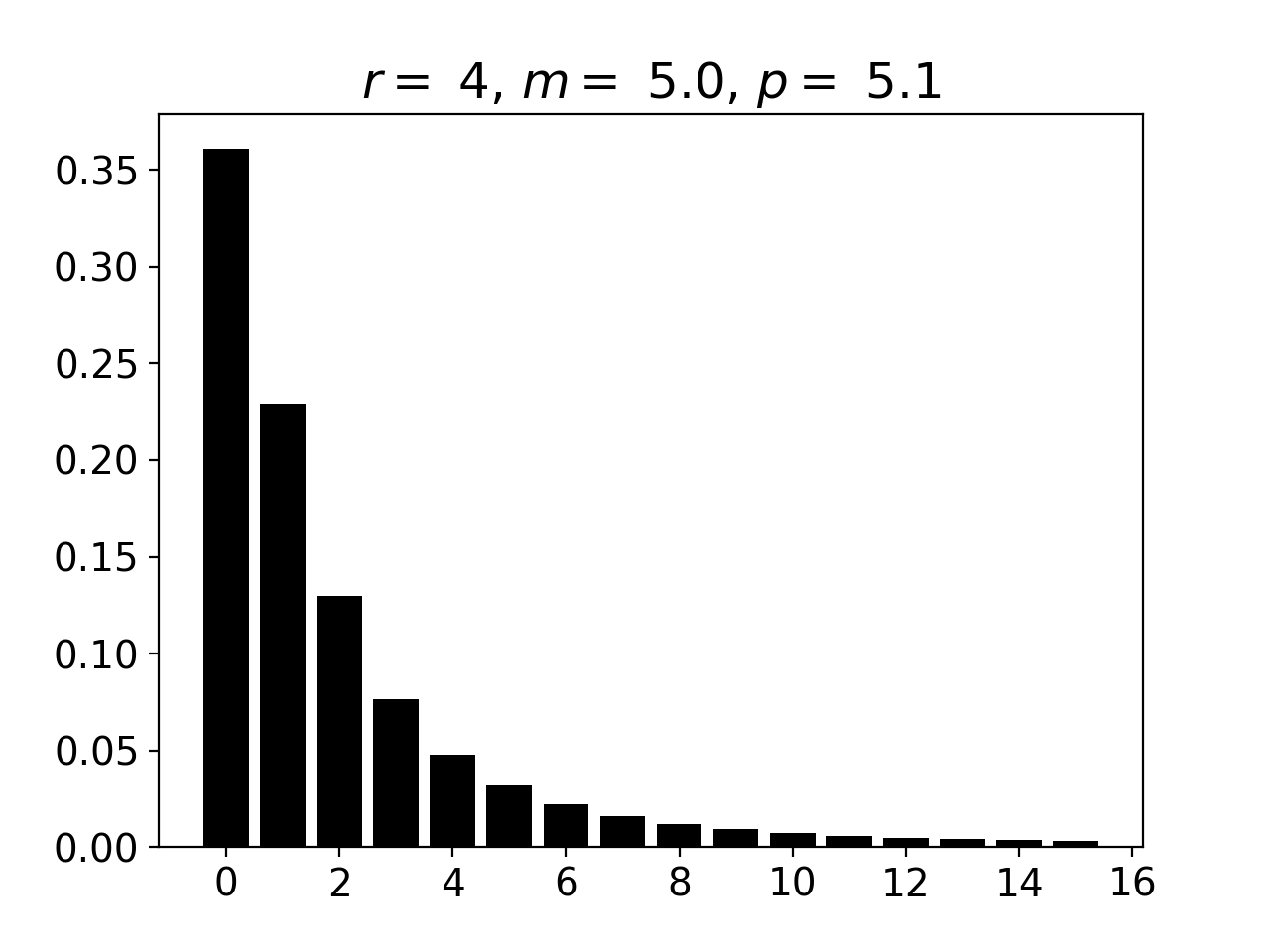}
\includegraphics[width=0.32\textwidth]{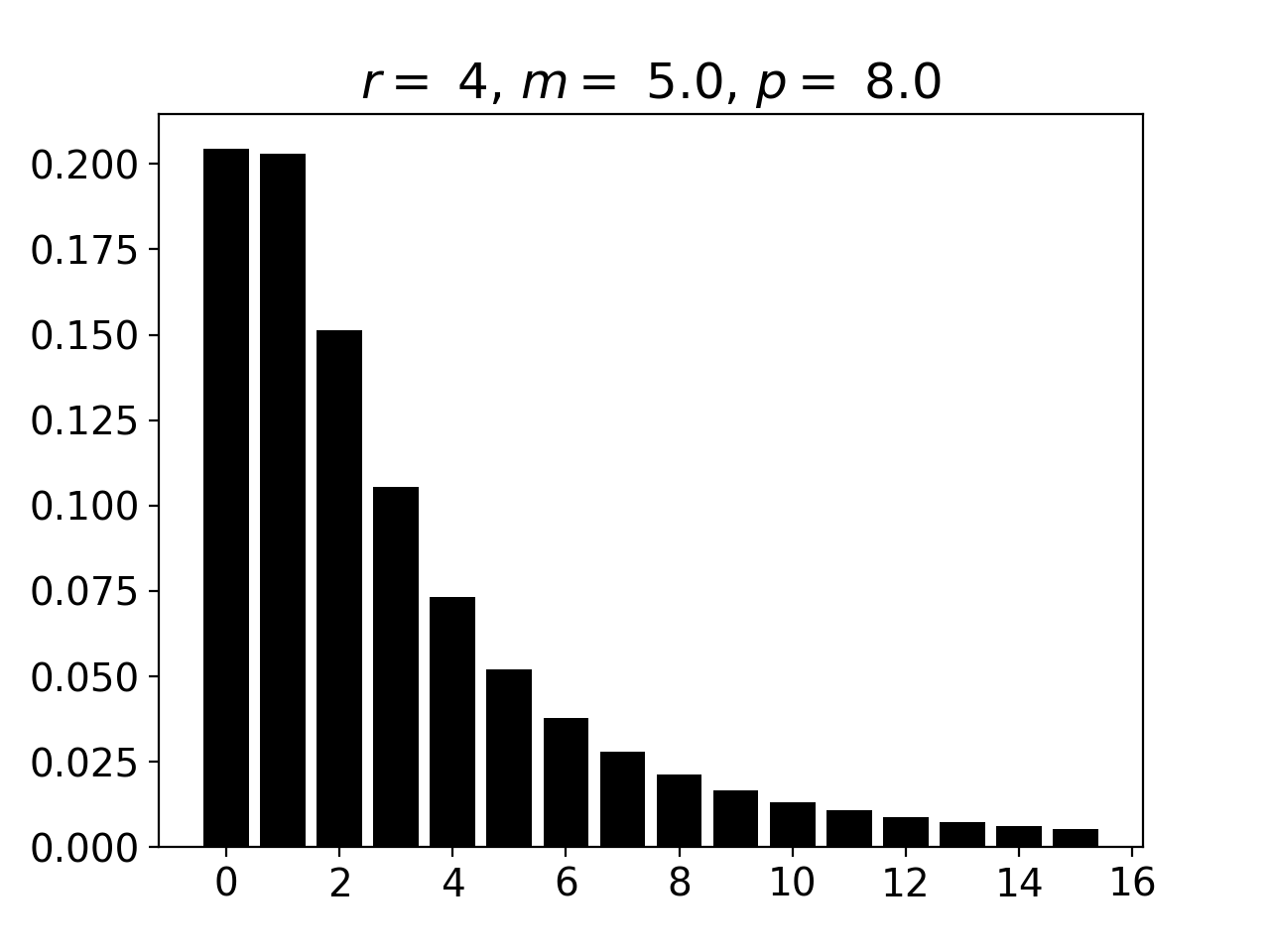}
\includegraphics[width=0.32\textwidth]{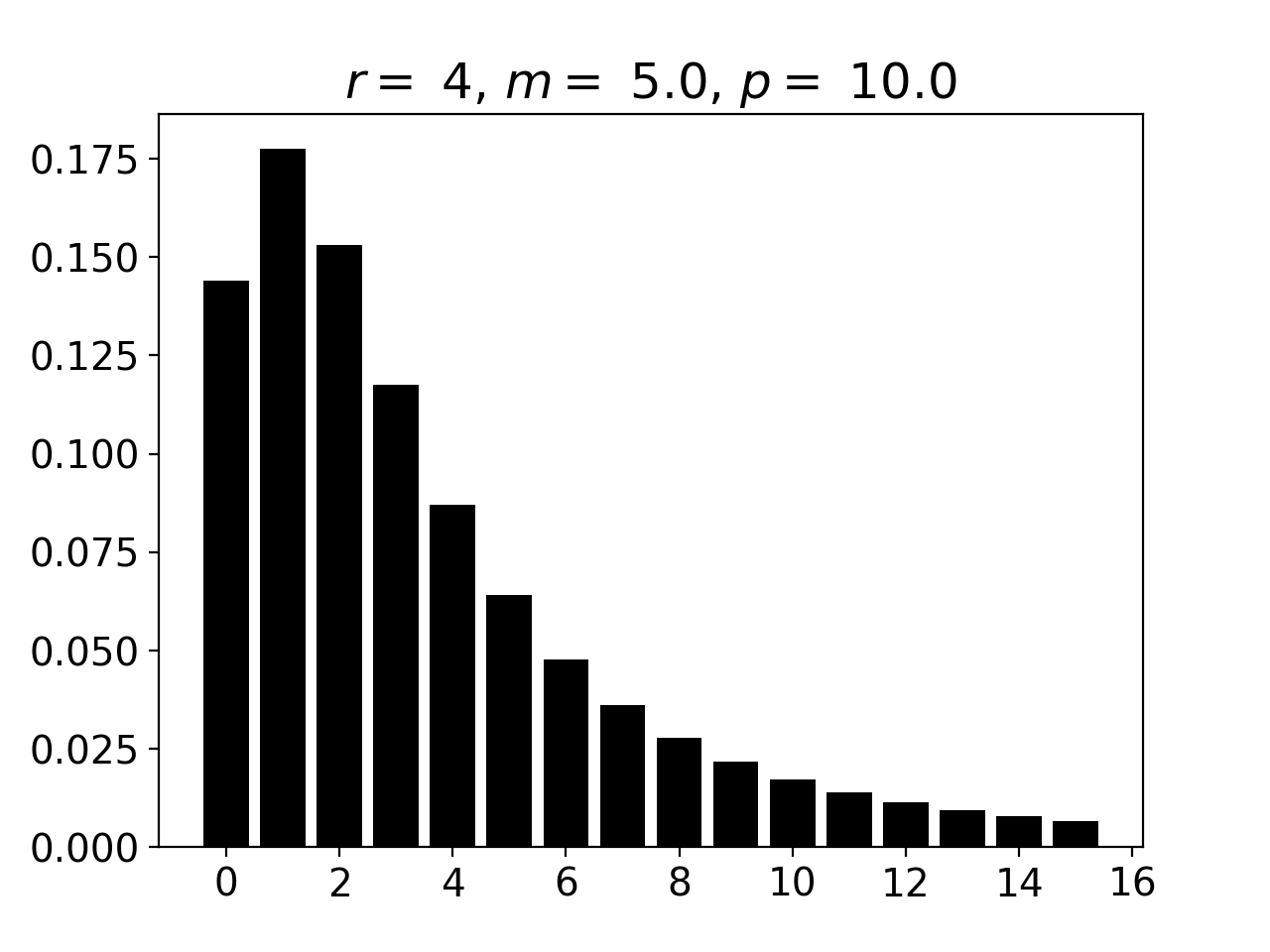}


\section{Data Applications}\label{s:data} 
Well-known data sets that are used very often for validation
and comparison reasons, concern automobile insurance claims because these
show heavy overdispersion and zero inflation. Other sets can found in
diverse fields such as marketing, biometry, health and social sciences. In
this section we investigate a number of these sets, refer to the best fitted
models that we could find in literature, and compare with our models. Within
our three classes we choose the one with the highest $p$-value of the
chi-square test, see Section \ref{ss:fitting}, where we computed the
distributions with variance functions up to power $r=9$.

\medskip\noindent 
We applied maximum likelihood for estimating the
parameters of our distributions, however, taking the mean parameter $m=%
\overline{x}$. Given the pair $\big(m,V(m)\big)$, we are able to compute the
probability mass function \eqref{e:pmf} numerically as function of the
parameter $p$ (for ABM and LMNS), or $p$ and $b$ (for LMS). Recall that the
data are $x=(n_0,\ldots,n_K)$. Then, the loglikelihood function is 
\begin{equation}  \label{e:ll}
\ell(p) = \log\prod_{k=0}^K f_{(m,p)}(k)^{n_k}= 
\sum_{k=0}^K n_k\big(\log(\mu_k) + k\psi(m) - \psi_1(m)\big),
\end{equation}
where $\mu_k, \psi(m)$, and $\psi_1(m)$ are functions of $p$. In case of the
LMS, the loglikelihood function is bivariate, $\ell(p,b)$. Maximizing the
loglikelihood function is done by a numerical optimization method.

\subsection{Data Set 1}
Insurance claims in Switserland in 1961 \citep{gossiaux1981}. We consider
the analyses of 
[1] \citep{willmot1987}, 
[2] \citep{gomez2011a}, and compare
with our models of 
[3] ABM($r=9$), 
[4] LMS ($r=3$), and 
[5] LMNS($r=1$). 
The data are clearly zero-inflated ($p_0^{\mathrm{emp}}=0.8653$), and
overdispersed (index of dispersion is $D=1.156$).

\begin{table}[H]
\caption{Claim data and fitted models}
\label{t:set1}
\medskip \centering
\begin{tabular}{rrrrrrr}
\# of claims & frequency & [1] & [2] & [3] & [4] & [5] \\ \hline
0 & 103704 & 103710.03 & 103347.35 & 103719.83 & 103718.88 & 103707.97 \\ 
1 & 14075 & 14054.65 & 14628.38 & 14016.51 & 14014.93 & 14060.87 \\ 
2 & 1766 & 1784.91 & 1682.27 & 1823.34 & 1827.88 & 1781.15 \\ 
3 & 255 & 254.49 & 175.79 & 250.35 & 249.38 & 252.84 \\ 
4 & 45 & 40.42 & 17.38 & 36.38 & 35.66 & 40.91 \\ 
5 & 6 & 6.94 & 1.65 & 5.55 & 5.31 & 7.40 \\ 
6 & 2 & 1.26 & 0.15 & 0.88 & 0.82 & 1.46 \\ \hline
$L$ &  & -54609.76 & -54659.61 & -54611.59 & -54612.03 & \textbf{-54609.75}
\\ 
$\chi^2$ &  & 0.7783 & 126.8 & 4.477 & 5.400 & 0.7432 \\ 
df &  & 3 & 4 & 3 & 2 & 3 \\ 
$p$-value &  & 0.8546 & 0.0 & 0.2143 & 0.0672 & \textbf{0.8630} \\ 
RMSE &  & 10.89 & 252.8 & 31.75 & 33.34 & \textbf{8.182}
\end{tabular}
\end{table}

\noindent 
Our LMNS model with $r=1$ gives the best fit, with the
Poisson-inverse Gaussian of \citet{willmot1987} as (almost) equal
competitor. The discrete Lindley distribution of \citet{gomez2011a} gives a
poor fit, as well our LMS models.

\subsection{Data Set 2}
Insurance claims in Zaire in 1974 \citep{gossiaux1981}. We consider the
analyses of 
[1] \citep{willmot1987}, 
[2] \citep{gomez2011b}, 
[3] \citep{bhati2019}, 
and compare with our models of
[4] ABM($r=9$), 
[5] LMS ($r=5$), and 
[6] LMNS($r=4$). 
The data are clearly zero-inflated 
($p_0^{\mathrm{emp}}=0.9298$), and overdispersed (index of dispersion is $D=1.417$).

\begin{table}[H]
\caption{Claim data and fitted models}
\label{t:set2}
\medskip \centering
\begin{tabular}{rrrrrrrr}
\# of claims & frequency & [1] & [2] & [3] & [4] & [5] & [6] \\ \hline
0 & 3719 & 3718.58 & 3719.06 & 3718.30 & 3718.98 & 3719.65 & 3718.83 \\ 
1 & 232 & 234.54 & 228.65 & 234.01 & 232.18 & 231.08 & 233.19 \\ 
2 & 38 & 34.86 & 41.85 & 36.09 & 37.29 & 37.57 & 36.30 \\ 
3 & 7 & 8.32 & 8.32 & 8.13 & 8.36 & 8.48 & 8.28 \\ 
4 & 3 & 2.45 & 1.68 & 2.26 & 2.22 & 2.25 & 2.30 \\ 
5 & 1 & 0.80 & 0.40 & 0.72 & 0.65 & 0.66 & 0.72 \\ \hline
$L$ &  & -1183.52 & -1183.97 & -1183.44 & -1183.37 & \textbf{-1183.36} & 
-1183.41 \\ 
$\chi^2$ &  & 0.5438 & 2.235 & 0.6240 & 0.4481 & 0.4555 & 0.3827 \\ 
df &  & 2 & 2 & 2 & 2 & 1 & 2 \\ 
$p$-value &  & 0.7619 & 0.3147 & 0.7320 & 0.7993 & 0.4997 & \textbf{0.8258}
\\ 
RMSE &  & 1.760 & 2.235 & 1.296 & \textbf{0.7212} & 0.8470 & 1.043
\end{tabular}
\end{table}

\noindent 
When we consider the $\chi^2$ criterion, we we find again the best
fit by an LMNS model ($r=4$), however the ABM and LMS models give better
(smaller) root mean squared errors. The Poisson-inverse Gaussian of 
\citet{willmot1987} and the geometric discrete Pareto of \citet{bhati2019}
are good competitors.

\subsection{Data Set 3}
Insurance claims in Germany in 1960 \citep{gossiaux1981}. We consider the
analyses of 
[1] \citep{willmot1987}, 
[2] \citep{gomez2011a}, 
[3] \citep{kokonendji2004a}, 
and compare with our models of 
[4] ABM($r=9$), 
[5] LMS ($r=3$), and 
[6] LMNS($r=1$). 
The data are clearly zero-inflated ($p_0^{%
\mathrm{emp}}=0.8729$), and overdispersed (index of dispersion is $D=1.136$).

\begin{table}[H]
\caption{Claim data and fitted models}
\label{t:set3}
\medskip \centering
\begin{tabular}{rrrrrrrr}
\# of claims & frequency & [1] & [2] & [3] & [4] & [5] & [6] \\ \hline
0 & 20592 & 20595.74 & 20544.79 & 20685.83 & 20596.75 & 20598.34 & 20595.56
\\ 
1 & 2651 & 2638.81 & 2720.36 & 2663.08 & 2633.91 & 2630.78 & 2639.47 \\ 
2 & 297 & 308.08 & 292.41 & 171.42 & 313.69 & 315.12 & 307.61 \\ 
3 & 41 & 39.68 & 28.55 & 55.00 & 38.81 & 38.97 & 39.50 \\ 
4 & 7 & 5.65 & 2.64 & 9.62 & 5.04 & 5.02 & 5.73 \\ 
5 & 0 & 0.87 & 0.24 & 3.24 & 0.68 & 0.67 & 0.93 \\ 
6 & 1 & 0.14 & 0.02 & 0.54 & 0.10 & 0.09 & 0.16 \\ \hline
$L$ &  & -10221.87 & -10228.45 & -10263.11 & -10222.51 & -10222.64 & 
\textbf{-10221.78} \\ 
$\chi^2$ &  & 0.7588 & 16.38 & 98.33 & 1.924 & 2.146 & 0.6649 \\ 
df &  & 2 & 3 & 2 & 2 & 1 & 2 \\ 
$p$-value &  & 0.6843 & $<0.001$ & 0.0 & 0.3821 & 0.1430 & \textbf{0.7172}
\\ 
RMSE &  & 6.442 & 32.15 & 59.68 & 9.282 & 10.60 & \textbf{6.136}
\end{tabular}
\end{table}

\noindent 
Our LMNS model with $r=1$ gives the best fit, with the
Poisson-inverse Gaussian of \citet{willmot1987} as (almost) equal
competitor. The reported discrete Lindley distribution of 
\citet{gomez2011a}, and the strict arcsine of \citet{kokonendji2004a} 
give poor fits.

\subsection{Data Set 4}
The number of European red mites on apple leaves \citep{bliss1953}. We
consider the analyses of 
[1] \citep{chakraborty2012}, 
[2] \citep{alamatsaz2016}, 
and compare with our models of 
[3] ABM($r=2$), 
[4] LMS ($r=1$), and 
[5] LMNS($r=9$). 
The data are clearly zero-inflated 
($p_0^{\mathrm{emp}}=0.4666$), and overdispersed (index of dispersion is $D=1.983$).

\begin{table}[H]
\caption{Red mites data and fitted models}
\label{t:set4}
\medskip \centering
\begin{tabular}{rrrrrrr}
\# of red mites & frequency & [1] & [2] & [3] & [4] & [5] \\ \hline
0 & 70 & 69.67 & 71.09 & 68.85 & 69.25 & 67.89 \\ 
1 & 38 & 37.49 & 32.08 & 38.90 & 38.20 & 40.51 \\ 
2 & 17 & 20.02 & 20.76 & 20.04 & 20.04 & 20.19 \\ 
3 & 10 & 10.67 & 12.88 & 10.35 & 10.50 & 10.00 \\ 
4 & 9 & 5.69 & 7.25 & 5.43 & 5.55 & 5.11 \\ 
5 & 3 & 3.03 & 3.60 & 2.90 & 2.69 & 2.71 \\ 
6 & 2 & 1.61 & 1.54 & 1.57 & 1.59 & 1.49 \\ 
7 & 1 & 0.86 & 0.56 & 0.86 & 0.86 & 0.84 \\ 
8 & 0 & 0.96 &  & 0.48 & 0.47 & 0.49 \\ \hline
$L$ &  & -222.44 & \textbf{-221.24} & -222.75 & -222.59 & -223.29 \\ 
$\chi^2$ &  & 2.896 & 2.868 & 3.461 & 3.180 & 4.483 \\ 
df &  & 5 & 5 & 5 & 4 & 5 \\ 
$p$-value &  & 0.7160 & \textbf{0.7204} & 0.6293 & 0.5281 & 0.4821 \\ 
RMSE &  & \textbf{1.563} & 2.635 & 1.656 & 1.578 & 2.018
\end{tabular}
\end{table}

\noindent 
The best (comparable) fits were given by the discrete Gamma 
of \citet{chakraborty2012}, and discrete Rayleigh of \citet{alamatsaz2016} (the
latter less good for the RMSE criterion). Our ABM model with $r=2$ comes
close.

\subsection{Data Set 5}
The number of accidents experienced by machinists \citep{bliss1953}. We
consider the analyse of 
[1] \citep{bhati2019}, 
and compare with our models of 
[2] ABM($r=9$), 
[3] LMS ($r=4$), and 
[4] LMNS($r=3$). 
The data are
clearly zero-inflated ($p_0^{\mathrm{emp}}=0.7150$), and overdispersed
(index of dispersion is $D=2.092$).

\begin{table}[H]
\caption{Accidents data and fitted models}
\label{t:set5}
\medskip \centering
\begin{tabular}{rrrrrr}
\# of accidents & frequency & [1] & [2] & [3] & [4] \\ \hline
0 & 296 & 296.60 & 295.91 & 296.44 & 295.30 \\ 
1 & 74 & 72.34 & 74.37 & 73.61 & 76.23 \\ 
2 & 26 & 25.48 & 24.80 & 24.83 & 24.20 \\ 
3 & 8 & 10.47 & 9.90 & 10.00 & 9.37 \\ 
4 & 4 & 4.68 & 4.43 & 4.50 & 4.18 \\ 
5 & 4 & 2.21 & 2.14 & 2.18 & 2.06 \\ 
6 & 1 &  & 1.10 & 1.11 & 1.09 \\ 
7 & 0 & 2.21 & 0.58 & 0.59 & 0.61 \\ 
8 & 1 &  & 0.32 & 0.32 & 0.36 \\ \hline
$L$ &  & -381.82 & -381.80 & \textbf{-381.78} & -381.95 \\ 
$\chi^2$ &  & 2.205 & 0.8985 & 0.9239 & 0.7534 \\ 
df &  & 3 & 3 & 2 & 3 \\ 
$p$-value &  & 0.820 & 0.8258 & 0.6300 & \textbf{0.8606} \\ 
RMSE &  & 1.373 & \textbf{1.035} & 1.060 & 1.297
\end{tabular}
\end{table}

\noindent 
All four models give competitive fits with respect to the RMSE
criterion. The LMS model shows worse in terms of $p$-value.

\subsection{Data Set 6}
The number of hospitalizations per family per year \citep{klugman1998}. We
consider the analyse of 
[1] \citep{gomez2011b}, 
and compare with our models of 
[2] ABM($r=9$), 
[3] LMS ($r=3$), and 
[4] LMNS($r=1$). 
The data are
clearly zero-inflated ($p_0^{\mathrm{emp}}=0.9094$), and overdispersed
(index of dispersion is $D=1.075$).

\begin{table}[H]
\caption{Hospitalization data and fitted models}
\label{t:set6}
\medskip \centering
\begin{tabular}{rrrrrr}
\# of hospitalizations & frequency & [1] & [2] & [3] & [4] \\ \hline
0 & 2659 & 2659.02 & 2659.03 & 2659.03 & 2658.95 \\ 
1 & 244 & 243.79 & 243.80 & 243.78 & 244.05 \\ 
2 & 19 & 19.52 & 19.47 & 19.50 & 19.22 \\ 
3 & 2 & 1.54 & 1.56 & 1.55 & 1.61 \\ 
4+ & 0 & 0.11 & 0.13 & 0.13 & 0.15 \\ \hline
$L$ &  & \textbf{-969.06} & \textbf{-969.06} & \textbf{-969.06} & -969.07 \\ 
$\chi^2$ &  & 0.07649 & 0.0634 & 0.2786 & 0.0320 \\ 
df &  & 1 & 1 & 1 & 1 \\ 
$p$-value &  & 0.7821 & 0.8011 & 0.5976 & \textbf{0.8581} \\ 
RMSE &  & 0.3278 & 0.3060 & 0.3205 & \textbf{0.2153}
\end{tabular}
\end{table}

\noindent 
All four models give competitive fits. Slightly the best would be
the LMNS model.

\subsection{Discussion}
The remarks on the six worked out data fits describe typically our findings.
We have executed an extensive numerical study on many more data sets, that
were considered in literature, see our report \citet{barlev2020b} which is
available on the arXiv. However, when we considered any study in the
literature on fitting count data, we noticed that it applied the proposed
model to just a few data sets, and then it gave good fits. When we ran the
model to other data sets, the picture could be changed drastically. On the
other hand, concerning our framework, the general observation is that in all
cases one of our models gives the best fit, or is competitive with the best
fit. Unfortunately, we have not yet discovered what underlying property of
the data makes that an ABM, or a LMS, or a LMNS will give the best fit.

\section{Conclusion}
This paper is an accompanying study of the computational aspects, and of the
practical usage of the framework of exponential dispersion models that we
have developed in \citet{barlev2020a}. These models act as alternatives not
only to classic distributions such as Poisson, generalized Poisson, negative
binomial, discrete Lindley, and Poisson-inverse Gaussian, but also to
recently propoped models, such as discrete Gamma, discrete Rayleigh, new
logarithmic, geometric discrete Pareto, and exponentiated discrete Lindley,
to name a few.

We have explained how our distributions are computed, where our starting
point are the solutions to integral equations involving the variance
functions of the mean parameterization of associated natural exponential
families. This enables us to use our framework for statistical modeling
purposes. Specifically, we considered the application of modeling
overdispersed, zero-inflated count data that occur in insurance, health
economics, incident reporting, and many others. We showed that our models
perform good or best in all cases when compared to best models reported in
literature.

\bigskip \noindent 
\textbf{Acknowledgements}. Shaul Bar-Lev was partially
supported in this research by the Netherlands Organization for Scientific
Research (NWO), project number 040.11.711.

\end{document}